\documentclass{article}

\usepackage{arxiv}

\usepackage[utf8]{inputenc} 
\usepackage[T1]{fontenc}    
\usepackage{hyperref}       
\usepackage{url}            
\usepackage{booktabs}       
\usepackage{amsfonts}       
\usepackage{nicefrac}       
\usepackage{microtype}      
\usepackage{graphicx}
\usepackage[numbers]{natbib}
\usepackage{doi}

\usepackage[T1]{fontenc}
\usepackage{hyperref}
\hypersetup{
colorlinks = true, 
linkcolor=black, 
citecolor=black, 
urlcolor=black 
}
\usepackage{verbatim}
\usepackage{tabularx} 
\usepackage{booktabs} 
\usepackage{makecell} 
\usepackage{xspace}
\usepackage{paralist}
\usepackage{stfloats}
\usepackage[table]{xcolor}
\usepackage{multirow}
\usepackage{subcaption}
\usepackage{cleveref}
\Crefname{figure}{Fig.}{Figs.}
\usepackage{url}
\usepackage[breakable]{tcolorbox}
\usepackage{svg}
\svgsetup{
inkscapepath=inkscape,
inkscapelatex=false
}

\def\BibTeX{{\rm B\kern-.05em{\sc i\kern-.025em b}\kern-.08em
T\kern-.1667em\lower.7ex\hbox{E}\kern-.125emX}}

\makeatletter
\newcommand{\linebreakand}{%
\end{@IEEEauthorhalign}
\hfill\mbox{}\par
\mbox{}\hfill\begin{@IEEEauthorhalign}
}
\makeatother

\newcommand{\finding}[1]{\begin{tcolorbox}[colframe=black, width=1\linewidth, colback=gray!15, left=2pt,right=2pt,top=2pt,bottom=2pt, breakable]#1\end{tcolorbox}}

\newcommand{\ENT}{\thead {ENT}} 
\newcommand{\DG}{\thead {DG}}
\newcommand{\MSP}{\thead {MSP}}
\newcommand{\PCS}{\thead {PCS}}
\newcommand{\ECEJ}{\thead{ECE \\ {\scriptsize (Jaccard)}}}
\newcommand{\MCEJ}{\thead{MCE \\ {\scriptsize (Jaccard)}}}

\newcommand{\deepseekLong}{DeepSeek-VL2\xspace}
\newcommand{\blipLong}{InstructBLIP\xspace}
\newcommand{\lavaLong}{LLaVA-1.6\xspace}
\newcommand{\qwenLong}{QWen2.5-VL\xspace}
\newcommand{\deepseek}{DeepSeek\xspace}
\newcommand{\blip}{BLIP\xspace}
\newcommand{\lava}{LLaVA\xspace}
\newcommand{\qwen}{QWen\xspace}
\newcommand{\TN}{\ensuremath{\mathit{TN}}\xspace}

\newcommand{\MCE}{\ensuremath{\mathit{MCE}}\xspace}

\newcommand{\trashcan}{TrashCan1.0\xspace}
\newcommand{\seaclear}{SeaClear\xspace}
\newcommand{\trashcans}{TrashCan\xspace} 

\definecolor{lightgray}{HTML}{A2A2A2}

\title{Assessing Vision–Language Models for Perception in Autonomous Underwater Robotic Software}

\author{ 
    \begin{minipage}[t]{0.32\textwidth}
    \centering
        {\hspace{1mm}Muhammad Yousaf} \\
    	{\normalfont Simula Research Laboratory and \\
    	Oslo Metropolitan University\\
    	Oslo, Norway} \\
    	\texttt{muhammady@simula.no} \\
    \end{minipage}
    \begin{minipage}[t]{0.32\textwidth}
        \centering
        {\hspace{1mm}Aitor Arrieta} \\
    	{\normalfont Mondragon University \\
    	Mondragon, Spain} \\
    	\texttt{aarrieta@mondragon.edu} \\
    \end{minipage}
    \begin{minipage}[t]{0.32\textwidth}
        \centering
        {\hspace{1mm}Shaukat Ali} \\
    	{\normalfont Simula Research Laboratory \\
    	Oslo, Norway} \\
    	\texttt{shaukat@simula.no} \\
    \end{minipage}
	\AND
    \begin{minipage}[t]{0.32\textwidth}
        \centering
        {\hspace{1mm}Paolo Arcaini} \\
    	{\normalfont National Institute of Informatics \\
    	Tokyo, Japan} \\
    	\texttt{arcaini@nii.ac.jp} \\
    \end{minipage}
    \begin{minipage}[t]{0.32\textwidth}
        \centering
        {\hspace{1mm}Shuai Wang} \\
    	{\normalfont Det norske Veritas (DNV) \\
    	Oslo, Norway} \\
    	\texttt{shuai.wang@dnv.com} \\
    \end{minipage}
}

\date{}

\renewcommand{\headeright}{ }
\renewcommand{\undertitle}{}
\renewcommand{\shorttitle}{\raggedright Assessing Vision–Language Models for Perception in Autonomous Underwater Robotic Software}

\hypersetup{
pdftitle={A template for the arxiv style},
pdfsubject={q-bio.NC, q-bio.QM},
pdfauthor={David S.~Hippocampus, Elias D.~Striatum},
pdfkeywords={First keyword, Second keyword, More},
}

\begin{document}
\maketitle

\begin{abstract}
Autonomous Underwater Robots (AURs) operate in challenging underwater environments, including low visibility and harsh water conditions. Such conditions present challenges for software engineers developing perception modules for the AUR software. To successfully carry out these tasks, deep learning has been incorporated into the AUR software to support its operations. However, the unique challenges of underwater environments pose difficulties for deep learning models, which often rely on labeled data that is scarce and noisy. This may undermine the trustworthiness of AUR software that relies on perception modules. Vision-Language Models (VLMs) offer promising solutions for AUR software as they generalize to unseen objects and remain robust in noisy conditions by inferring information from contextual cues. Despite this potential, their performance and uncertainty in underwater environments remain understudied from a software engineering perspective. Motivated by the needs of an industrial partner in assurance and risk management for maritime systems to assess the potential use of VLMs in this context, we present an empirical evaluation of VLM-based perception modules within the AUR software. We assess their ability to detect underwater trash by computing performance, uncertainty, and their relationship, to enable software engineers to select appropriate VLMs for their AUR software.
\end{abstract}


\section{Introduction}\label{sec:introduction}
Autonomous Underwater Robots (AURs) are deployed across a wide range of applications, including industrial tasks (e.g., monitoring and repairing subsea production systems, or collecting underwater debris) and scientific research (e.g., studying marine life~\cite{saoud2023mars}). Operating underwater presents many challenges, including poor visibility and unstable environmental conditions. These conditions pose significant challenges for the development and quality assurance of reliable AUR software in industrial contexts. Increasingly, machine learning techniques, such as deep learning, are being applied to enhance AUR operations, including perception, object detection, route planning, and data analysis~\cite{UOD_survey_2025,lightweight_debris_detection_2026}. However, these models often struggle due to challenging underwater conditions and the scarcity of labeled data, thereby affecting the dependability of the industrial AUR software.

Vision-Language Models (VLMs)~\cite{vlm_comprehensive_survey2025} offer advanced capabilities to support AUR software, especially for perception, due to their generalization abilities, multi-modal data understanding, and strong reasoning skills. Examples include the identification of unexpected objects not present in labeled datasets, recognition of objects in challenging environments (e.g., low-visibility environments, common in underwater), and support for advanced data analysis~\cite{Alawode2025AquaticCLIP,vla_dualBrain_2025}.

VLMs typically serve as backbone controllers for Visual Language Action (VLA) models~\cite{ma2024survey}, a new generation of embodied AI techniques, including software modules for general-purpose robotic manipulators (e.g., GR00T-N1~\cite{bjorck2025gr00t}) and underwater robots~\cite{gu2025usim}. As backbone controllers, if these VLMs are not reliable software components, the VLA will produce low-quality outputs. Therefore, from a software testing perspective, it is important to test VLMs, in isolation, for their ability to identify and list visible objects underwater before deployment in AUR software, since unreliable VLM components will compromise the safety of the AUR as a whole.

As part of the European project InnoGuard~\cite{innoguard}, whose objective is to develop methods to ensure the dependability of autonomous cyber-physical systems (ACPSs), we aim to address the assurance and risk-management needs of our industrial partner DNV AS, in the era of large AI models. DNV is a global assurance and risk management company with a strong focus on maritime autonomous cyber-physical systems. To this end, to enable the industrial use of VLMs in AUR, we present an empirical evaluation of VLM effectiveness for underwater trash collection, providing evidence for their use as components of perception modules in AUR software. Specifically, we assess their ability to detect and classify objects, analyze uncertainty, and examine the relationship between performance and uncertainty. We selected four VLMs (\blip\cite{Li2022BLIP}, \lava\cite{liu2024Lava}, \deepseek\cite{Liu2024DeepSeekV3}, and \qwen\cite{bai2023qwen}) and two underwater datasets (\trashcan\cite{trashcan_arxiv}, \seaclear\cite{seaclear_nature}), and defined metrics to evaluate performance and uncertainty.

Overall, \blip and \deepseek perform the best, with \blip being slightly better. For Trash and Object classes, \blip and \deepseek outperform \qwen and \lava, while the performance of all models is poor for Vegetation and Animal classes. \lava is very confident and highly certain, but poorly calibrated, indicating overconfidence and making it unreliable for use in industrial AUR software's perception module. \blip is a little less confident and more uncertain than \lava, but it is the best-calibrated model, making it a reliable choice for deployment in industrial AUR software. These results suggest that high performance aligns more with good calibration than with high confidence or low uncertainty, highlighting the importance of prioritizing both performance and calibration when selecting VLMs for integration into AUR software.

To summarize, our key contributions are:
\begin{inparaenum}[1)]
\item an empirical evaluation of four open-source VLMs to assess their ability to detect objects in underwater images and determine their suitability for use in AUR software,
\item a study of the relationship between performance, confidence, uncertainty, and calibration to identify which factors are most important for selecting VLMs as part of perception modules in AUR software. 
\end{inparaenum}


\section{Industrial Context}\label{sec:overview}
This work is part of the EU project InnoGuard~\cite{innoguard}, which aims to develop methods to ensure the quality of Autonomous Cyber-Physical Systems (ACPSs), e.g., humanoid robots, autonomous vessels, and AURs. The project involves several academic institutions and industrial partners focused on ACPS quality assurance. The work presented in this paper focuses on one particular industrial partner, DNV AS, whose main business is assurance and risk management for reliable and trustworthy industrial systems in multiple domains, including maritime systems, which is the focus of this paper.

As AI foundation models increasingly attract industrial attention, the assurance process should evaluate whether these models, including VLMs, can be reliably integrated into various maritime applications, including supporting the operations of AURs. Potential applications include using VLMs as components of AUR software for complex tasks (e.g., perception) that are traditionally handled by alternative methods, as well as integrating VLMs into digital twins of AURs to enable more advanced analyses and simulations in real-time.

Given the safety-critical nature of the maritime domain, ensuring the quality and safety of AURs is essential. VLMs are relatively new models, and our industrial partner currently lacks evidence on how they can be reliably integrated into AURs for tasks such as underwater object identification, where failures can have severe operational consequences. This creates a clear need for empirical evaluations to determine whether VLMs can be trusted in real-world maritime applications, included as components of ACPS software.

Moreover, our industrial partner is interested in understanding which VLMs can meet established quality and safety standards, align with existing software engineering workflows, and comply with regulatory requirements in the domains where these systems are deployed. This industrial context motivates systematic empirical evaluations that test VLMs for AUR software, providing evidence to guide both industrial stakeholders and software engineers in selecting reliable VLMs for ACPS software. As a first step, this work presents a systematic empirical study of VLMs as software components in AURs, aimed at assessing their reliability and generating evidence to support industrial adoption in the maritime domain.

\section{Experiment Design} \label{sec:experimentdesign}
We present the empirical evaluation design. Sect.~\ref{subsec:overview} presents the setup, Sect.~\ref{subsec:RQs} the research questions, and Sect.~\ref{subsec:benchmarks} the datasets. Sect.~\ref{subsec:experimental_settings} describes the models, configurations, and prompts, while Sect.~\ref{subsec:EvaluationMetrics} defines the evaluation metrics. The replication package is available online~\cite{SupplementaryMaterialICSE2026}. \looseness=-1

\subsection{Overall Empirical Evaluation Setup}\label{subsec:overview}
The overall objective of our empirical evaluation is to assess whether VLMs can provide a viable solution to support AUR's trash collection operations implemented in its software. In particular, we aim to examine their potential to identify trash items, including their confidence, especially for unknown trash items that deep learning models might miss.

\begin{figure}[!t]
\centering
\includegraphics[width=0.8\linewidth]{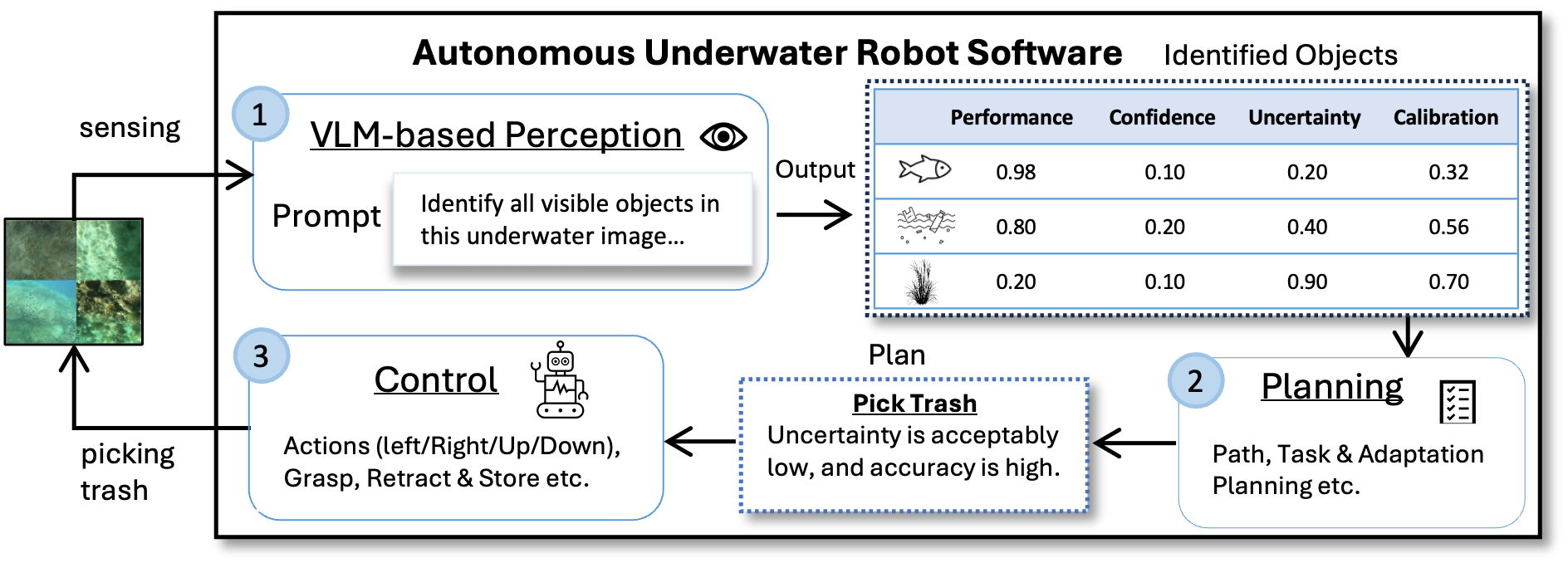}
\caption{Overall Architecture of the AUR Software}
\label{fig:overview}
\end{figure}

Fig.~\ref{fig:overview} presents the overall architecture of a typical VLM that we use, focusing on analyzing underwater images collected by onboard cameras of an AUR.
An image, along with a text instruction (e.g., ``Identify and list visible objects in this underwater image'') and the set of target labels for classification (e.g., ``trash''), are provided in a prompt to the VLM. The image is processed through a \emph{vision encoder}, while the text instruction is handled by a \emph{text encoder}. The vision encoder extracts visual features, which are projected into a shared embedding space by the \emph{vision-language projector} and then combined with textual embeddings for a \emph{large language model}. The integrated representation enables the VLM to perform classification tasks for underwater images in our context. We assess classification performance using standard metrics (e.g., F1 score) and quantify uncertainty using class probabilities produced from a softmax output layer. The output includes predicted labels and confidence scores for each identified class, which are used to quantify uncertainty with various probability-based metrics (e.g., probability confidence score).

Using this overall architecture, we planned our empirical evaluation, which is described in this section, while results and analyses are presented in Sect.~\ref{sec:results}.

\subsection{Research Questions}\label{subsec:RQs}
We defined the following research questions:
\begin{compactdesc}
\item \textbf{RQ1 (Performance) How well do VLMs classify underwater images?}
This RQ helps assess the performance of VLMs for underwater object classification tasks, enabling evaluation of their suitability for trash detection in AURs.
\item \textbf{RQ2 (Uncertainty Quantification) How uncertain are VLMs when classifying underwater images?}
This RQ helps us evaluate uncertainty in VLMs while classifying underwater objects, which is essential for safe and effective trash detection by AURs.
\item \textbf{RQ3 (Performance--Uncertainty Relationship) What is the relationship between performance and uncertainty in underwater image classification?}
Studying the performance–uncertainty relationship helps understand whether VLMs can support AURs in making reliable and safe decisions when detecting trash. For instance, if the VLM performs well at detecting trash but has high uncertainty, it indicates that the VLM is underconfident.
\end{compactdesc}

\subsection{Benchmark Datasets}\label{subsec:benchmarks}
To assess the performance and uncertainty of VLMs in underwater multi-label classification, we employ two benchmark datasets: \trashcan~\cite{trashcan_arxiv} and \seaclear~\cite{seaclear_nature}. Even though the datasets provide more detailed instance and material-level labels, our study focuses on the general categorization by original dataset, which closely aligns with AURs tasked with trash detection and collection, i.e., \emph{Animal}, \emph{Vegetation}, \emph{Object}, and \emph{Trash}. However, the original \emph{Bio} class from source datasets was separated into \emph{Animal} and \emph{Vegetation}, while man-made objects and trash classes were retained as \emph{Trash}. It also supports zero-shot evaluation of VLMs, produces improved uncertainty estimates, and reduces the need for fine-tuning on domain-specific datasets, where foundational models often struggle in low-visibility underwater conditions. This allows evaluating the generalization ability of VLMs in detecting and recognizing unseen underwater objects.


The \trashcan dataset contains different classes, including marine animals (e.g., fish, crabs), vegetation (e.g., algae, plants), man-made objects (e.g., bottles, nets), and trash.

The \seaclear dataset contains both deep and shallow underwater scenes. Its labels include marine litter, man-made objects, equipment, and marine life. Compared to \trashcan, it emphasizes plastic debris and containers, making it useful for testing VLMs performance and uncertainty in scenarios that contain mixed clutter from natural and artificial categories.

Table~\ref{tab:dataset_comparison} summarizes the two datasets used in this empirical study in terms of the number of images and the percentage of images that contain each of the four classes.
\begin{table}[!t]
\caption{Characteristics of the Selected Datasets. \#Images is the number of images in each dataset; Class (\%) is the percentage of images that contain an instance of a give class.}
\label{tab:dataset_comparison}
\centering
\setlength{\tabcolsep}{12pt}
\begin{tabular}{l c c c c c}
\toprule
\textbf{Dataset} & \textbf{\#Images} & \multicolumn{4}{c}{\textbf{Class (\%)}} \\
\cmidrule(lr){3-6}
& & \textbf{Object} & \textbf{Trash} & \textbf{Animal} & \textbf{Vegetation} \\
\midrule
\texttt{\trashcans} & 7,212 & 82\% & 67\% & 19\% & 7\% \\
\texttt{\seaclear} & 8,610 & 93\% & 77\% & 39\% & 5\% \\
\bottomrule
\end{tabular}

\end{table}

\subsection{Experimental Setting}\label{subsec:experimental_settings}
This sub-section discusses the selected VLMs, including their settings, prompt design, and execution environment.

\paragraph{\textbf{Subject VLMs}}

We selected the following four recent open-source VLMs, released in 2024 or later~\cite{vlm_survey_2025}, that represent current multimodal architectures capable of text–image understanding relevant to our classification task and also provide the token-level logits access required for uncertainty quantification computation:
\begin{inparaenum}[(i)]
\item \deepseekLong,
\item \blipLong
\item \lava, and 
\item \qwenLong. 
\end{inparaenum}
Table~\ref{tab:vlm_comparison_split} presents the key characteristics of the selected VLMs, including their visual encoders and other notable features. It also reports the ID that will be used in the paper to identify them. These VLMS can perform a wide variety of vision tasks, such as image captioning, scene understanding, and video reasoning, thereby better understanding complex underwater environments. These differences can affect the model's performance, as models with strong grounding and reasoning capabilities (e.g., \qwen and \deepseek) tend to achieve better results. 
\begin{table}[!t]
\centering
\caption{Comparison of Selected Vision-Language Models.}
\label{tab:vlm_comparison_split}
\centering
\setlength{\tabcolsep}{20pt}
\begin{tabular}{l c p{160pt}}
\toprule
\textbf{Model} (\textbf{ID}) & \textbf{Visual Encoder} & \textbf{Special Features} \\
\midrule
\blipLong (\blip) & ViT & Visual Dialogue, Visual Question Answering, Captioning, Retrieval \\
\addlinespace
\lavaLong (\lava) & CLIP ViT-L/14 & High-Resolution Visual Grounding \\
\addlinespace
\deepseekLong (\deepseek) & SigLIP, SAMB & Video Analysis, Scene Understanding, 3D Reconstruction \\
\addlinespace
\qwenLong (\qwen) & Redesigned ViT & Visual Dialogue, Visual Question Answering, Captioning, Image reasoning \\
\bottomrule
\end{tabular}
\end{table}
To ensure a fair evaluation across models, the above-selected VLMs have a similar parameter count of 7B each. Different parameter counts affect the model's complexity and performance, ranging from less to more powerful behavior.
Evaluation metrics were computed per image and aggregated at the dataset level, reporting summary statistics including mean, standard deviation, minimum, and maximum values.
For each input image, a model generated textual predictions, token-level probability distributions, which served as the basis for evaluating performance and uncertainty metrics. To obtain deterministic predictions suitable for our classification task, all VLMs were executed with the same configurations and a zero temperature value for deterministic outputs.


\paragraph{\textbf{Prompt Design}}
A single domain-dependent, zero-shot, and instruction-style same prompt was used across all VLMs and datasets to elicit multi-label outputs with semantic consistency.
Table~\ref{tab:prompt_template} reports the prompt template and an example output. This design enforces structured outputs, while allowing open-vocabulary reasoning, enabling fair cross-model comparison. Each model's generated text was post-processed using the exact consistent label mapping to four classes to compute per-image performance metrics, including Precision, Recall, F1-scores, and Jaccard similarity index.
\begin{table}[!t]
\caption{Prompt template and example output}
\label{tab:prompt_template}

\setlength{\tabcolsep}{20pt}
\begin{tabular}{p{40pt}p{200pt}p{100pt}}
\toprule
\textbf{Component} & \textbf{Prompt Template} & \textbf{Example Output} \\
\midrule
\makecell[t]{System Message} & ``Identify and list all visible objects in this underwater image in four categories: (1) Animals, (2) Vegetation, (3) Objects, and (4) Trash. For each category, list the detected items with counts. If a category is definitely absent, write ‘None'.'' & (same across all VLMs) \\
\midrule
{Animal} & Animals: [list detected animals with counts] & Animals: 1 fish \\
\addlinespace
{Vegetation} & Vegetation: [list detected plants with counts] & Vegetation: 1 plant \\
\addlinespace
{Object} & Objects: [list man-made objects with counts] & Objects: 1 plastic bottle \\
\addlinespace
{Trash} & Trash: [list trash items with counts] & Trash: 1 piece of trash \\
\midrule
\makecell[t]{Final\\Labels} & Predicted category & \{`trash', `object', `vegetation', `animal'\} \\
\bottomrule
\end{tabular}
\end{table}

\paragraph{\textbf{Execution Environment}}
All experiments were conducted in Python and PyTorch on a workstation equipped with an NVIDIA RTX-5090 GPU. Each VLM was evaluated in \emph{zero-shot inference mode}, without any additional training or fine-tuning, to ensure a fair comparison across models.

\subsection{Evaluation Metrics} \label{subsec:EvaluationMetrics}
We here introduce the evaluation metrics used to answer the RQs. We present evaluation metrics for RQ1 and RQ2. For RQ3, which studies the relationship between performance and uncertainty, we use the same metrics as for RQ1 and RQ2. 

\subsubsection{RQ1 -- Evaluation Metrics for Performance}\label{subsubsec:accuracymetrics}

To assess the performance of the studied VLMs, we selected metrics relevant to our task, i.e., a multi-label classification problem. Unlike binary classification, where a single F1 score can be applied, we used two variants of F1 suitable for multi-label classification: F1 (Macro) and F1 (Micro). These metrics allow us to evaluate the VLMs' performance from different perspectives. F1 (Macro) treats each class equally and helps assess the overall performance of the VLMs across all classes together (i.e., Animal, Vegetation, Object, and Trash), while F1 (Micro) treats each sample equally, thus giving more weight to classes with more samples. In our case, F1 (Micro) allows us to assess the overall VLM performance irrespective of class imbalance in the datasets. In addition, we selected the Jaccard index, which provides a complementary perspective to F1 on VLM performance. Specifically, it evaluates performance by measuring the overlap between predicted and true class labels. Since our problem is a multi-label classification task, we used Jaccard (Macro) and Jaccard (Micro), which function similarly to the Macro and Micro F1 scores. In addition, we evaluated the performance of VLMs per-class (e.g., Trash), using standard F1 scores and the Jaccard index.

\subsubsection{RQ2 -- Evaluation Metrics for Uncertainty Quantification} \label{subsec:unertainty_metrics}
We employ probability-based uncertainty quantification metrics to quantify predictive confidence from token-level probabilities, thereby capturing model-prediction uncertainty. In particular, we assessed the uncertainty of the studied VLMs for our task from three perspectives. First, we evaluated their confidence in their stated classification certainty. Second, we assessed uncertainty by quantifying the spread of predicted probabilities across classes; the higher the spread, the higher the uncertainty. Third, we examined calibration errors to evaluate the reliability of the VLMs confidence estimates.

\paragraph{\textbf{Confidence Metrics}}
VLMs generate a sequence of textual tokens corresponding to multiple predicted labels for a given input image $I$. At each token position, the model applies a softmax function to its logits to produce a probability distribution over the vocabulary. To quantify model's confidence at each token, we use the {\it Max Softmax Probability} and the {\it Probability Confidence Score}, which select the highest predicted probability within the distribution. The higher the confidence values, the lower the VLM's uncertainty about its task. These metrics operate on the token-level probability distributions $\{p_1,$ $\ldots,$ $p_n\}$ predicted for each of \TN output tokens. The per-token confidence score is defined as:

\smallskip

\noindent\textit{(1) Max Softmax Probability (MSP):} 
It measures the model confidence and selects the highest softmax probability assigned to predicted token or output.
\begin{equation}
P_i = \max_j p_{i,j}
\label{eq:msp_token}
\end{equation}
where $p_{i,j}$ is the probability assigned to the $j$-th possible token at position $i$, and $P_i$ the model's confidence for that prediction.
\begin{equation}
\mathit{MSP}(I) = \frac{1}{\TN} \sum_{i=1}^{\TN} P_i
\label{eq:msp_average}
\end{equation}
Here, \TN is the total number of generated tokens for image $I$. Higher ${\mathit{MSP}}$ values indicate greater model confidence, while lower values correspond to less confident predictions.

\smallskip

\noindent\textit{(2) Probability Confidence Score (PCS):}
It measures the margin between the highest and second-highest probabilities within each token's probability distribution:
\begin{equation}
\mathit{Margin}_i = P_{i,(1)} - P_{i,(2)}
\label{eq:pcs_token}
\end{equation}
where $P_{i,(1)}$ and $P_{i,(2)}$ denote the highest and second highest predicted probabilities for the $i$-th token, respectively. A large $\mathit{Margin}_i$ shows a strong preference for the top candidate, while a small value means ambiguity between token choices.

\begin{equation}
\mathit{PCS}(I) = \frac{1}{\TN} \sum_{i=1}^{\TN} \mathit{Margin}_i
\label{eq:pcs_average}
\end{equation}
where \TN is the total number of generated tokens for image $I$. 
Higher $\mathit{PCS}$ values shows greater model confidence, while lower values mean less confident predictions.

\paragraph{\textbf{Uncertainty Metrics}}

We use {\it Entropy} and {\it Deep Gini} as representative uncertainty measures, as both are computed directly from token‑level softmax distributions providing complementary perspectives for assessing the uncertainty of VLMs. Entropy quantifies overall dispersion, indicating how evenly probability mass is spread across tokens, whereas Deep Gini measures probability inequality, emphasizing how concentrated or diffuse the model's prediction is.

\smallskip

\noindent\textit{(1) Deep Gini (DG):}
It quantifies distributional sharpness. A uniform distribution (spread across many classes) corresponds to high uncertainty. The token-level DG is defined as:
\begin{equation}
\mathit{Impurity}_i = 1 - \sum_{j=1}^{N} p_{i,j}^2
\label{eq:gini_token}
\end{equation}
where $p_{i,j}$ is the probability assigned to the $j$-th possible token at position $i$. 
A low $\mathit{Impurity}$ in probability distribution reflects high uncertainty. The per-image DG uncertainty is the average across all generated tokens:
\begin{equation}
\mathit{DG}(I) = \frac{1}{\TN} \sum_{i=1}^{\TN} \mathit{Impurity}_i
\label{eq:gini_average}
\end{equation}
where \TN is defined as before. 
Higher $\mathit{DG}$ values indicate greater overall model uncertainty.

\smallskip

\noindent\textit{(2) Entropy (ENT):}
It captures the spread of probabilities, increasing when the model assigns similar probabilities across many classes. In information theory entropy is represented by symbol $H$. The token-level entropy is calculated as:
\begin{equation}
H_i = - \sum_{j=1}^{N} p_{i,j} \log p_{i,j}
\label{eq:entropy_token}
\end{equation}
where $p_{i,j}$ represents the probability of the $j$-th token at position $i$, and $N$ denotes the total number of possible tokens. The image-level entropy-based uncertainty is obtained by averaging over all tokens:
\begin{equation}
\mathit{ENT}(I) = \frac{1}{\TN} \sum_{i=1}^{\TN} H_i
\label{eq:entropy_average}
\end{equation}
where \TN is defined as before. 
Higher $\mathit{ENT}$ values correspond to greater uncertainty in the model's predictions.

\paragraph{\textbf{Calibration Metrics}}
We used calibration-based metrics to assess the alignment between predicted confidence and actual correctness, thereby quantifying the quality of models' confidence. Poor calibration (high error) indicates that the model's confidence estimates cannot be trusted. In particular, we used the {\it Expected Calibration Error} and the {\it Maximum Calibration Error}, where the former measures the average miscalibration across all predictions, whereas the latter captures the worst miscalibration. Together, they provide complementary assessments of VLM calibration errors.

\smallskip

\noindent\textit{(1) Expected Calibration Error (ECE):}
It measures average miscalibration across $M$ confidence bins where predictions are grouped into bins based on their confidence scores:
\begin{equation}
\mathit{ECE} = \sum_{m=1}^M \frac{|B_m|}{N} \, \big| F1(B_m) - MSP(B_m) \big|
\label{eq:ece_equation}
\end{equation}
where $\mathit{F1}(B_m)$ denotes the F1 (Micro) score, which measures the average performance of VLM across samples in bin $m$, a suitable metric for our multi-label classification task. $\mathit{MSP}(B_m)$ is their average predicted confidence (\ref{eq:msp_average}), $B_m$ is the samples in bin $m$ and $N$ is the total number of samples.

\smallskip

\noindent\textit{(2) Maximum Calibration Error (MCE):}
Highlights the worst-case bin-level deviation between model's confidence and overall performance (F1 (Micro) in our case):
\begin{equation}
\MCE = \max_{m \in \{1,\dots,M\}} \big| \mathit{F1}(B_m) - \mathit{MSP}(B_m) \big|
\label{eq:mce_equation}
\end{equation}
where $\mathit{F1}(B_m)$ denotes the F1 (Micro) score which measures the average performance, $\mathit{MSP}(B_m)$ is their average predicted confidence of a VLM and $B_m$ specifies the number of samples in bin $m$. A high \MCE means that at least one confidence interval has a large calibration error. A perfectly calibrated model would have $\MCE=0$.

\section{Results and Analysis} \label{sec:results}

\subsection{RQ1 -- Performance of VLMs}
RQ1 examines the performance of VLMs in a multi-label classification setting for underwater images, evaluated across two benchmark datasets, both individually and in aggregate. The analysis is conducted from two perspectives: overall performance across all four classes and per-class performance.

\subsubsection{Overall Performance}
\begin{figure}[!tb]
\centering %
\begin{subfigure}[b]{\columnwidth}
\centering
\includegraphics[width=0.6\linewidth]{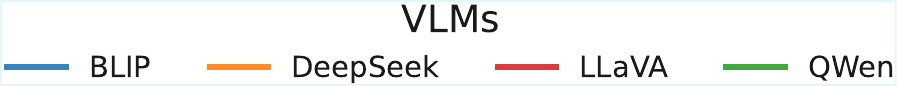}
\end{subfigure}
\begin{subfigure}[b]{0.325\columnwidth}
\centering
\includegraphics[width=\linewidth]{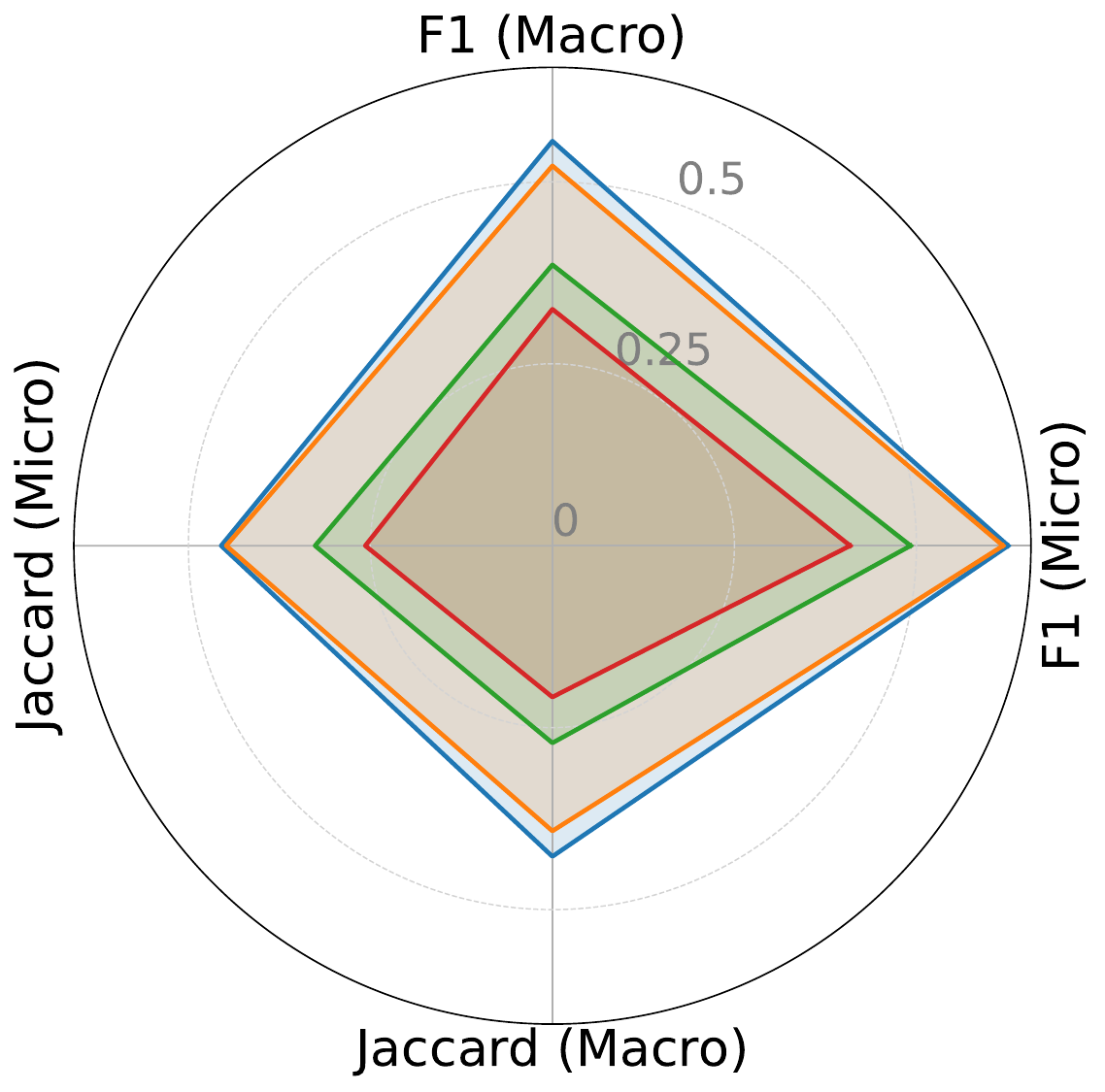}
\label{fig:RQ1_accuracy_metrics_radar_trash}
\caption{TrashCan1.0}
\end{subfigure}
\begin{subfigure}[b]{0.325\columnwidth}
\centering
\includegraphics[width=\linewidth]{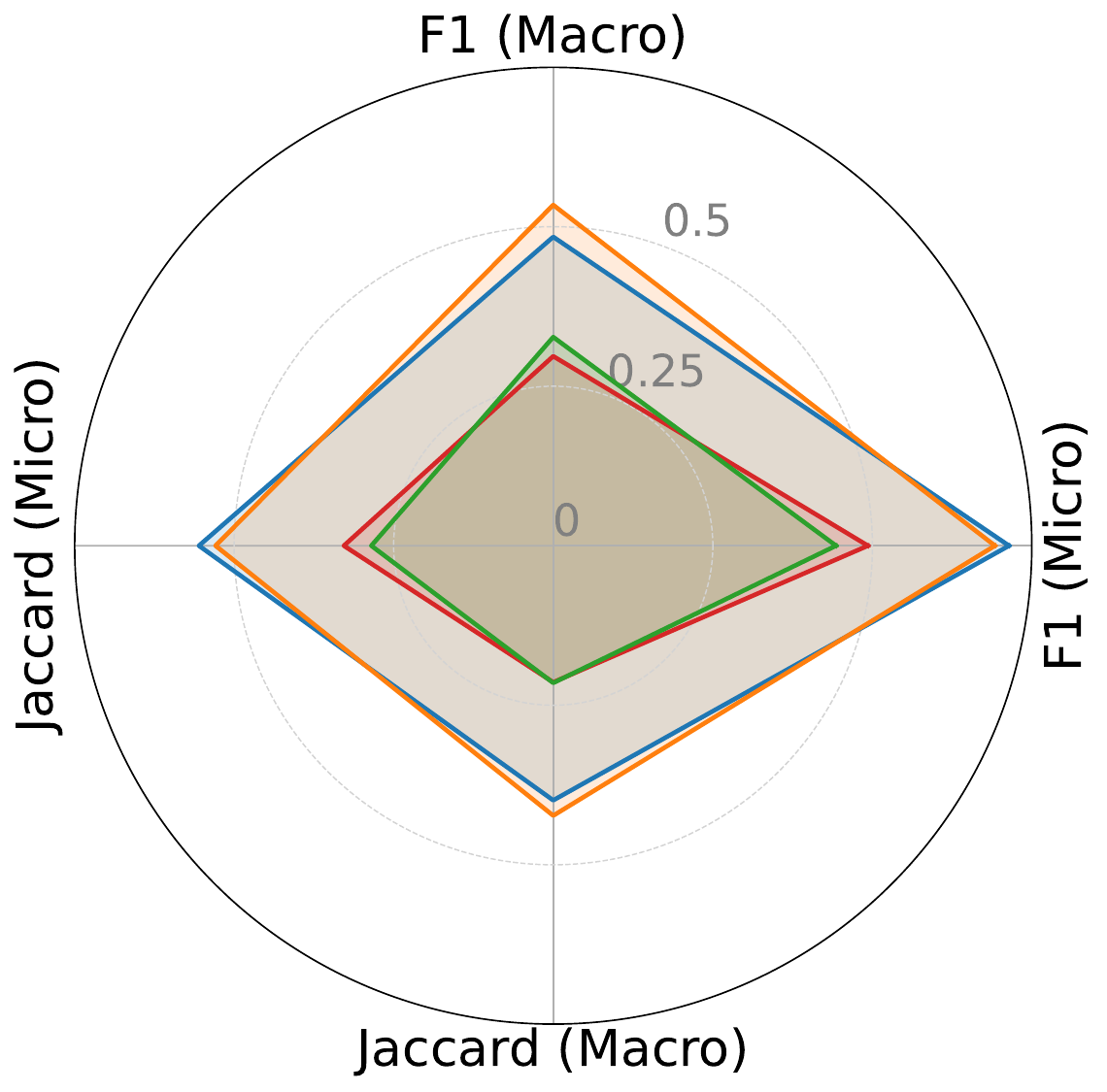}
\label{fig:RQ1_accuracy_metrics_radar_seaClear}
\caption{SeaClear}
\end{subfigure}
\begin{subfigure}[b]{0.325\columnwidth}
\centering
\includegraphics[width=\linewidth]{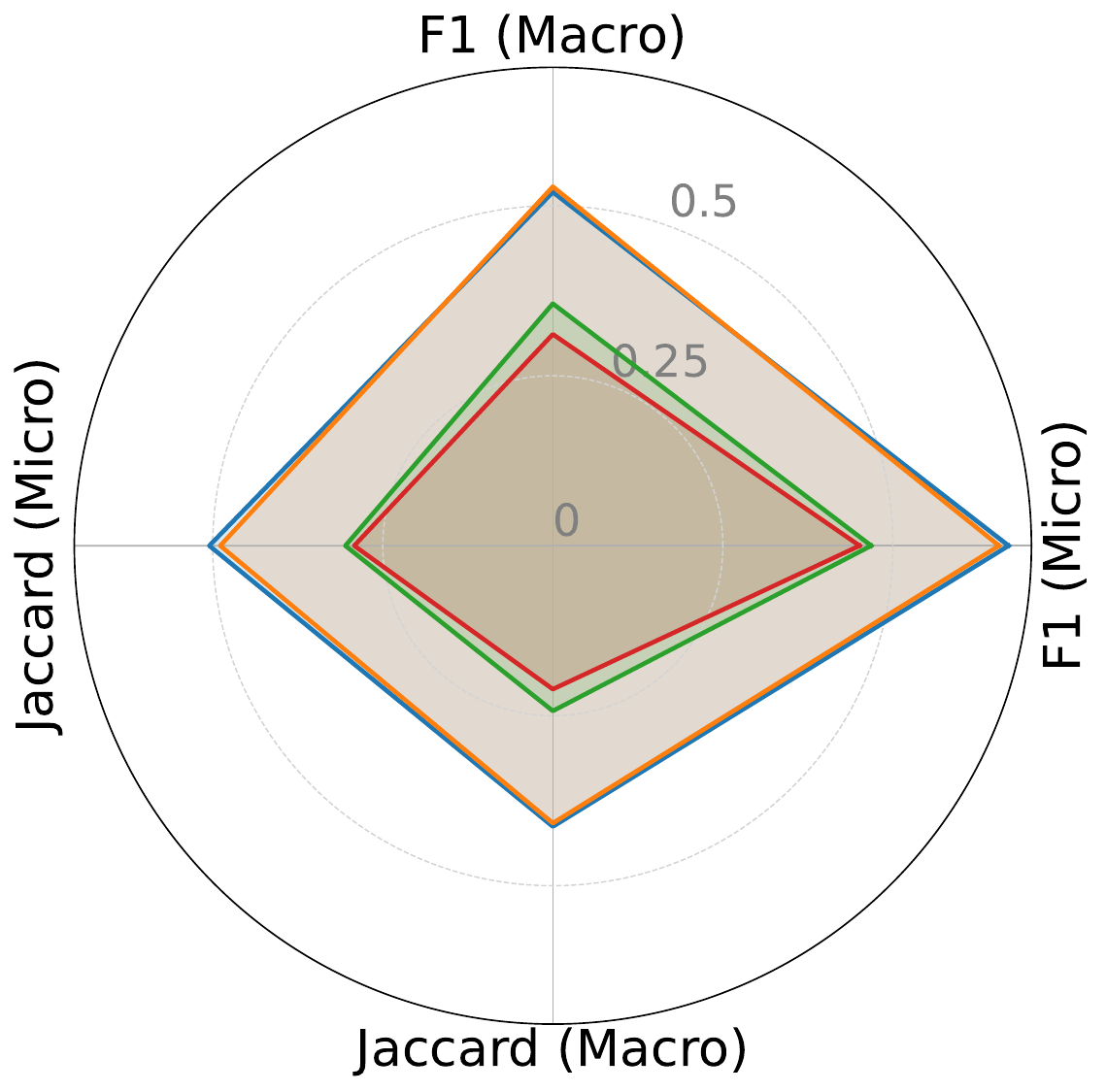}
\label{fig:RQ1_accuracy_metrics_radar_combined}
\caption{Aggregated}
\end{subfigure}
\caption{RQ1 -- Comparison of overall performance metrics for the VLMs across datasets}
\label{fig:RQ1_accuracy_metrics_radar}
\end{figure}

Table~\ref{tab:RQ1_accuracy_metrics_summary} and Fig.~\ref{fig:RQ1_accuracy_metrics_radar} present the overall results.
\begin{table}[!tb]
\caption{RQ1 -- Overall performance of VLMs across datasets. The darker the gray, the better the VLMs perform.}
\label{tab:RQ1_accuracy_metrics_summary}
\setlength{\tabcolsep}{5pt}

\begin{tabular}{lrrrr|rrrr|rrrr}
\toprule
& \multicolumn{4}{c}{TrashCan1.0} & \multicolumn{4}{c}{SeaClear} & \multicolumn{4}{c}{Aggregated} \\
\cmidrule(lr){2-5} \cmidrule(lr){6-9} \cmidrule(lr){10-13}
Model & \thead{F1 \\ {\scriptsize (Micro)}} & \thead{F1 \\ {\scriptsize (Macro)}} & \thead{Jaccard \\ {\scriptsize (Micro)}} & \thead{Jaccard \\ {\scriptsize (Macro)}} & \thead{F1 \\ {\scriptsize (Micro)}} & \thead{F1 \\ {\scriptsize (Macro)}} & \thead{Jaccard \\ {\scriptsize (Micro)}} & \thead{Jaccard \\ {\scriptsize (Macro)}} & \thead{F1 \\ {\scriptsize (Micro)}} & \thead{F1 \\ {\scriptsize (Macro)}} & \thead{Jaccard \\ {\scriptsize (Micro)}} & \thead{Jaccard \\ {\scriptsize (Macro)}} \\
\midrule
\blip & \cellcolor{lightgray!63}0.63 & \cellcolor{lightgray!56}0.56 & \cellcolor{lightgray!46}0.46 & \cellcolor{lightgray!43}0.43 & \cellcolor{lightgray!71}0.71 &\cellcolor{lightgray!48}0.48 & \cellcolor{lightgray!56}0.56 &\cellcolor{lightgray!40}0.40 & \cellcolor{lightgray!67}0.67 &\cellcolor{lightgray!52}0.52 & \cellcolor{lightgray!51}0.51 & \cellcolor{lightgray!41}0.41 \\
\deepseek & \cellcolor{lightgray!62}0.62 &\cellcolor{lightgray!52}0.52 &\cellcolor{lightgray!45}0.45 &\cellcolor{lightgray!39}0.39 &\cellcolor{lightgray!69}0.69 & \cellcolor{lightgray!53}0.53 &\cellcolor{lightgray!53}0.53 & \cellcolor{lightgray!42}0.42 &\cellcolor{lightgray!66}0.66 & \cellcolor{lightgray!53}0.53 &\cellcolor{lightgray!49}0.49 & \cellcolor{lightgray!41}0.41 \\
\lava & \cellcolor{lightgray!41}0.41 & \cellcolor{lightgray!33}0.33 & \cellcolor{lightgray!26}0.26 & \cellcolor{lightgray!21}0.21 &\cellcolor{lightgray!49}0.49 & \cellcolor{lightgray!30}0.30 &\cellcolor{lightgray!33}0.33 & \cellcolor{lightgray!21}0.21 & \cellcolor{lightgray!45}0.45 & \cellcolor{lightgray!31}0.31 & \cellcolor{lightgray!28}0.29 & \cellcolor{lightgray!21}0.21 \\
\qwen & \cellcolor{lightgray!49}0.49 &\cellcolor{lightgray!39}0.39 &\cellcolor{lightgray!33}0.33 &\cellcolor{lightgray!27}0.27 & \cellcolor{lightgray!44}0.44 &\cellcolor{lightgray!33}0.33 & \cellcolor{lightgray!28}0.28 & \cellcolor{lightgray!21}0.21 &\cellcolor{lightgray!47}0.47 &\cellcolor{lightgray!36}0.36 &\cellcolor{lightgray!31}0.31 &\cellcolor{lightgray!24}0.24 \\
\bottomrule
\end{tabular}

\end{table}
%
%
On \trashcan, \blip achieves the highest performance across all metrics among all models. For \seaclear, \blip performs best in terms of F1 (Micro) and Jaccard (Micro), while \deepseek outperforms it on the other two metrics; however, the differences from \blip are relatively small. In the aggregated dataset, \blip remains the best for F1 (Micro) and Jaccard (Micro), and ties with \deepseek for the Jaccard (Macro). For F1 (Macro) in the aggregated data, \deepseek slightly surpasses \blip (0.53 vs. 0.52), a minimal difference. Overall, \blip performs best across the models. The radar charts in Fig.~\ref{fig:RQ1_accuracy_metrics_radar} highlight similar trends, with \blip and \deepseek performing comparably and both outperforming \lava and \qwen. Considering the lowest performance, \lava generally ranks poorest, except for \seaclear, where \qwen has the lowest F1 (Micro) and Jaccard (Micro), though it ties with \lava on Jaccard (Macro).

\subsubsection{Per-Class Performance}
Table~\ref{tab:RQ1_accuracy_perclass_metrics} shows the per-class performance of VLMs in terms of F1 and Jaccard across the individual and aggregated datasets. 
\begin{table}[!tb]
\caption{RQ1 -- Per-Class performance of VLMs across datasets. The darker the gray, the better the VLMs perform. 
}
\label{tab:RQ1_accuracy_perclass_metrics}
\centering
\setlength{\tabcolsep}{10pt}
\begin{tabular}{llrrrrrr}
\toprule
&  & \multicolumn{2}{c}{TrashCan1.0} & \multicolumn{2}{c}{SeaClear} & \multicolumn{2}{c}{Aggregated} \\
\cmidrule(lr){3-4} \cmidrule(lr){5-6} \cmidrule(lr){7-8} 
Class & Model & F1 & Jaccard & F1 & Jaccard & F1 & Jaccard \\
\midrule
\multirow[c]{4}{*}{Trash} & \blip & \cellcolor{lightgray!72}0.72 & \cellcolor{lightgray!56}0.57 & \cellcolor{lightgray!80}0.80 & \cellcolor{lightgray!66}0.66 & \cellcolor{lightgray!76}0.76 & \cellcolor{lightgray!62}0.62 \\
& DeepSeek &\cellcolor{lightgray!68}0.68 &\cellcolor{lightgray!51}0.51 &\cellcolor{lightgray!71}0.71 &\cellcolor{lightgray!55}0.55 &\cellcolor{lightgray!69}0.69 &\cellcolor{lightgray!53}0.53 \\
& \lava &\cellcolor{lightgray!31}0.31 &\cellcolor{lightgray!18}0.18 &\cellcolor{lightgray!32}0.32 &\cellcolor{lightgray!19}0.19 &\cellcolor{lightgray!32}0.32 &\cellcolor{lightgray!19}0.19 \\
& \qwen & \cellcolor{lightgray!22}0.22 & \cellcolor{lightgray!12}0.12 & \cellcolor{lightgray!31}0.31 & \cellcolor{lightgray!18}0.18 & \cellcolor{lightgray!26}0.26 & \cellcolor{lightgray!15}0.15 \\
\cmidrule{1-8}
\multirow[c]{4}{*}{Object} & \blip & \cellcolor{lightgray!87}0.87 & \cellcolor{lightgray!76}0.76 & \cellcolor{lightgray!88}0.88 & \cellcolor{lightgray!79}0.79 & \cellcolor{lightgray!87}0.87 & \cellcolor{lightgray!78}0.78 \\
& \deepseek &\cellcolor{lightgray!84}0.84 &\cellcolor{lightgray!72}0.72 & \cellcolor{lightgray!88}0.88 & \cellcolor{lightgray!79}0.79 &\cellcolor{lightgray!86}0.86 &\cellcolor{lightgray!76}0.76 \\
& \lava & \cellcolor{lightgray!56}0.57 & \cellcolor{lightgray!40}0.40 &\cellcolor{lightgray!76}0.76 &\cellcolor{lightgray!61}0.61 & \cellcolor{lightgray!66}0.66 & \cellcolor{lightgray!50}0.50 \\
& \qwen &\cellcolor{lightgray!71}0.71 &\cellcolor{lightgray!55}0.55 & \cellcolor{lightgray!63}0.63 & \cellcolor{lightgray!46}0.46 &\cellcolor{lightgray!67}0.67 &\cellcolor{lightgray!51}0.51 \\
\cmidrule{1-8}
\multirow[c]{4}{*}{Animal} & \blip & \cellcolor{lightgray!33}0.33 & \cellcolor{lightgray!20}0.20 &\cellcolor{lightgray!22}0.22 &\cellcolor{lightgray!12}0.12 &\cellcolor{lightgray!28}0.28 &\cellcolor{lightgray!16}0.16 \\
& \deepseek &\cellcolor{lightgray!34}0.34 &\cellcolor{lightgray!21}0.21 & \cellcolor{lightgray!49}0.49 & \cellcolor{lightgray!33}0.33 &\cellcolor{lightgray!42}0.42 &\cellcolor{lightgray!27}0.27 \\
& \lava &\cellcolor{lightgray!37}0.37 &\cellcolor{lightgray!23}0.23 & \cellcolor{lightgray!3}0.03 & \cellcolor{lightgray!1}0.01 & \cellcolor{lightgray!20}0.20 & \cellcolor{lightgray!12}0.12 \\
& \qwen & \cellcolor{lightgray!55}0.55 & \cellcolor{lightgray!38}0.38 &\cellcolor{lightgray!32}0.32 &\cellcolor{lightgray!19}0.19 & \cellcolor{lightgray!43}0.43 & \cellcolor{lightgray!28}0.28 \\
\cmidrule{1-8}
\multirow[c]{4}{*}{Vegetation} & \blip & \cellcolor{lightgray!31}0.31 & \cellcolor{lightgray!18}0.18 & \cellcolor{lightgray!4}0.04 & \cellcolor{lightgray!2}0.02 & \cellcolor{lightgray!17}0.17 & \cellcolor{lightgray!10}0.10 \\
& \deepseek &\cellcolor{lightgray!23}0.23 &\cellcolor{lightgray!13}0.13 &\cellcolor{lightgray!5}0.05 &\cellcolor{lightgray!3}0.03 &\cellcolor{lightgray!14}0.14 &\cellcolor{lightgray!8}0.08 \\
& \lava &\cellcolor{lightgray!5}0.05 & \cellcolor{lightgray!3}0.03 & \cellcolor{lightgray!8}0.08 & \cellcolor{lightgray!4}0.04 & \cellcolor{lightgray!6}0.06 & \cellcolor{lightgray!3}0.03 \\
& \qwen & \cellcolor{lightgray!7}0.07 &\cellcolor{lightgray!4}0.04 &\cellcolor{lightgray!5}0.05 &\cellcolor{lightgray!3}0.03 & \cellcolor{lightgray!6}0.06 & \cellcolor{lightgray!3}0.03 \\
\bottomrule
\end{tabular}

\end{table}
Since we are assessing per-class performance, the Micro and Macro versions of F1 and Jaccard are not applicable.
We have the following observations:
\begin{inparaenum}[(1)]
\item \textit{Trash:} \blip is the best-performing model, with minor differences from \deepseek, while the differences with \qwen and \lava are larger;
\item \textit{Object:} \blip is again the best-performing model, with minor differences from \deepseek. \qwen and \lava generally remain the worst models, although their differences from the best model are not as large as those observed for Trash; 
\item \textit{Animal:} \qwen is the best for \trashcan and in the aggregated dataset. \lava is the worst for \seaclear and the aggregated dataset, and \blip is the worst for \trashcan. Here, even though some models perform better than others, the overall performance of all VLMs is low. The highest value is 0.55 for F1 for \qwen in \trashcan, which indicates that, overall, the performance of VLMs for Animal is not good;
\item \textit{Vegetation:} \blip is the best for \trashcan and the aggregated dataset, while \lava is the best for \seaclear. For \seaclear, \blip has the worst performance, while in the aggregated dataset, both \lava and \qwen are the worst. For \trashcan, \lava is the worst performer in both F1 and Jaccard. Generally, the performance values for all metrics for Vegetation are quite low. For example, the highest F1 score is 0.31 for \blip on \trashcan. Thus, even though some models perform better than others, the overall performance for VLMs remains low for the Vegetation class.
\end{inparaenum}



\finding{\textbf{RQ1:} Overall, \blip and \deepseek show similar performance in classifying underwater images, with \blip performing slightly better. Considering per-class performance, \blip and \deepseek perform similarly for Trash and Object classes, outperforming \qwen and \lava, whereas for Vegetation and Animal, all models perform poorly.}

\subsection{RQ2 -- VLM Uncertainty Quantification}
RQ2 examines VLMs' uncertainty. We quantify overall uncertainty across all classes and per-class uncertainty using confidence, uncertainty, and calibration metrics.

\subsubsection{Overall Uncertainty Quantification}
Table~\ref{tab:RQ2_token-level_probability_summary} summarizes the overall results, without distinguising between the four classes.
%
\begin{table*}[!t]
\caption{RQ2 -- Uncertainty quantification results for VLMs across datasets, both individually and in aggregate. The darker the gray, the better the VLMs perform on uncertainty metrics. $\uparrow$ indicates that higher values correspond to better uncertainty, whereas $\downarrow$ indicates that lower values correspond to better uncertainty.}
\label{tab:RQ2_token-level_probability_summary}
\setlength{\tabcolsep}{4pt}
\resizebox{\textwidth}{!}{
\begin{tabular}{lrrrrrr|rrrrrr|rrrrrrrrrrrr}
\toprule
& \multicolumn{6}{c}{TrashCan1.0} & \multicolumn{6}{c}{SeaClear} & \multicolumn{6}{c}{Aggregated} \\
\cmidrule(lr){2-7} \cmidrule(lr){8-13} \cmidrule(lr){14-19}
& \multicolumn{2}{c}{Confidence $\uparrow$} & \multicolumn{2}{c}{Uncertainty $\downarrow$} & \multicolumn{2}{c}{Calibration $\downarrow$} & \multicolumn{2}{c}{Confidence $\uparrow$} & \multicolumn{2}{c}{Uncertainty $\downarrow$} & \multicolumn{2}{c}{Calibration $\downarrow$} & \multicolumn{2}{c}{Confidence $\uparrow$} & \multicolumn{2}{c}{Uncertainty $\downarrow$} & \multicolumn{2}{c}{Calibration $\downarrow$} \\
\cmidrule(lr){2-3} \cmidrule(lr){4-5} \cmidrule(lr){6-7} \cmidrule(lr){8-9} \cmidrule(lr){10-11} \cmidrule(lr){12-13} \cmidrule(lr){14-15} \cmidrule(lr){16-17} \cmidrule(lr){18-19}

Model & \MSP & \PCS & \ENT & \DG & \ECEJ & \MCEJ & 
\MSP & \PCS & \ENT & \DG & \ECEJ & \MCEJ & 
\MSP & \PCS & \ENT & \DG & \ECEJ & \MCEJ \\
\midrule
\blip & \cellcolor{lightgray!76}0.76 & \cellcolor{lightgray!66}0.66 & \cellcolor{lightgray!100}1.06 & \cellcolor{lightgray!32}0.32 & \cellcolor{lightgray!28}0.29 & \cellcolor{lightgray!34}0.34 & \cellcolor{lightgray!78}0.78 & \cellcolor{lightgray!71}0.71 & \cellcolor{lightgray!94}0.94 &\cellcolor{lightgray!28}0.29 & \cellcolor{lightgray!21}0.21 & \cellcolor{lightgray!28}0.29 & \cellcolor{lightgray!77}0.77 & \cellcolor{lightgray!69}0.69 & \cellcolor{lightgray!100}1.00 & \cellcolor{lightgray!31}0.31 & \cellcolor{lightgray!25}0.25 & \cellcolor{lightgray!32}0.32 \\
\deepseek &\cellcolor{lightgray!77}0.77 &\cellcolor{lightgray!71}0.71 &\cellcolor{lightgray!84}0.84 &\cellcolor{lightgray!30}0.30 &\cellcolor{lightgray!32}0.32 &\cellcolor{lightgray!49}0.49 &\cellcolor{lightgray!77}0.77 &\cellcolor{lightgray!72}0.72 &\cellcolor{lightgray!83}0.83 & \cellcolor{lightgray!31}0.31 &\cellcolor{lightgray!25}0.25 &\cellcolor{lightgray!44}0.44 & \cellcolor{lightgray!77}0.77 &\cellcolor{lightgray!72}0.72 &\cellcolor{lightgray!84}0.84 & \cellcolor{lightgray!31}0.31 &\cellcolor{lightgray!28}0.29 &\cellcolor{lightgray!46}0.46 \\
\lava & \cellcolor{lightgray!89}0.89 & \cellcolor{lightgray!87}0.87 & \cellcolor{lightgray!43}0.43 & \cellcolor{lightgray!14}0.14 & \cellcolor{lightgray!62}0.62 & \cellcolor{lightgray!84}0.84 & \cellcolor{lightgray!87}0.87 & \cellcolor{lightgray!84}0.84 & \cellcolor{lightgray!50}0.50 & \cellcolor{lightgray!17}0.17 & \cellcolor{lightgray!53}0.53 & \cellcolor{lightgray!84}0.84 & \cellcolor{lightgray!88}0.88 & \cellcolor{lightgray!85}0.85 & \cellcolor{lightgray!47}0.47 & \cellcolor{lightgray!16}0.16 & \cellcolor{lightgray!57}0.58 & \cellcolor{lightgray!84}0.84 \\
\qwen &\cellcolor{lightgray!78}0.78 &\cellcolor{lightgray!74}0.74 &\cellcolor{lightgray!89}0.89 &\cellcolor{lightgray!28}0.29 &\cellcolor{lightgray!44}0.44 &\cellcolor{lightgray!53}0.53 & \cellcolor{lightgray!78}0.78 &\cellcolor{lightgray!73}0.73 &\cellcolor{lightgray!86}0.86 &\cellcolor{lightgray!28}0.29 &\cellcolor{lightgray!51}0.51 &\cellcolor{lightgray!87}0.87 &\cellcolor{lightgray!78}0.78 &\cellcolor{lightgray!74}0.74 &\cellcolor{lightgray!88}0.88 &\cellcolor{lightgray!28}0.29 &\cellcolor{lightgray!48}0.48 &\cellcolor{lightgray!70}0.70 \\
\bottomrule
\end{tabular}
}
\end{table*}
For both individual and aggregated datasets, \lava is the most confident and certain model, yet it is the least well-calibrated, suggesting that it is overconfident. Instead, \blip is generally the least confident and relatively more uncertain than the rest of the VLMs, with a few exceptions: for DG in \seaclear where \deepseek is slightly more uncertain than \blip; for MSP in \seaclear, \blip ties with \qwen; and for MSP and DG in the aggregated dataset, \blip ties with \deepseek. However, even in these cases where \blip is the least confident and most uncertain, the differences compared to \lava (i.e., the most confident and certain) are not large. Consistently, \blip is the best-calibrated model. These results suggest that \blip is cautious in its classifications and its probability estimates are reliable, making it suitable for critical applications.

\subsubsection{Per-Class Uncertainty Quantification}
Results of per-class uncertainty are reported in Table~\ref{tab:RQ2_per_class_uq_summary}.
\begin{table}
\caption{RQ2 -- Per-Class Uncertainty Metrics for VLMs across datasets. The darker the gray, the better the VLMs perform on uncertainty metrics. $\uparrow$ indicates that higher values correspond to better uncertainty, whereas $\downarrow$ indicates that lower values correspond to better uncertainty.}
\label{tab:RQ2_per_class_uq_summary}
\centering
\setlength{\tabcolsep}{4pt}
\begin{tabular}{llrrrr|rrrr|rrrr}
\toprule
&  & \multicolumn{4}{c}{TrashCan1.0} & \multicolumn{4}{c}{SeaClear} & \multicolumn{4}{c}{Aggregated} \\
\cmidrule(lr){3-6} \cmidrule(lr){7-10} \cmidrule(lr){11-14}
& & \multicolumn{2}{c}{Confidence $\uparrow$} & \multicolumn{2}{c}{Uncertainty $\downarrow$} & \multicolumn{2}{c}{Confidence $\uparrow$} & \multicolumn{2}{c}{Uncertainty $\downarrow$} & \multicolumn{2}{c}{Confidence $\uparrow$} & \multicolumn{2}{c}{Uncertainty $\downarrow$} \\
\cmidrule(lr){3-4} \cmidrule(lr){5-6} \cmidrule(lr){7-8} \cmidrule(lr){9-10} \cmidrule(lr){11-12} \cmidrule(lr){13-14}
Class & Model & \MSP & \PCS & \ENT & \DG & \MSP & \PCS & \ENT & \DG & \MSP & \PCS & \ENT & \DG \\
\midrule
\multirow[c]{4}{*}{\rotatebox{90}{Trash}} 
& \blip & \cellcolor{lightgray!76}0.76 & \cellcolor{lightgray!67}0.67 & \cellcolor{lightgray!100}1.05 & \cellcolor{lightgray!32}0.32 &\cellcolor{lightgray!79}0.79 & \cellcolor{lightgray!71}0.71 & \cellcolor{lightgray!91}0.91 &\cellcolor{lightgray!28}0.28 &\cellcolor{lightgray!78}0.78 & \cellcolor{lightgray!69}0.69 & \cellcolor{lightgray!98}0.98 & \cellcolor{lightgray!30}0.30 \\
& \deepseek &\cellcolor{lightgray!78}0.78 &\cellcolor{lightgray!71}0.71 &\cellcolor{lightgray!84}0.84 &\cellcolor{lightgray!30}0.30 & \cellcolor{lightgray!77}0.77 & \cellcolor{lightgray!71}0.71 &\cellcolor{lightgray!83}0.83 & \cellcolor{lightgray!31}0.31 & \cellcolor{lightgray!77}0.77 &\cellcolor{lightgray!71}0.71 &\cellcolor{lightgray!84}0.84 & \cellcolor{lightgray!30}0.30 \\
& \lava & \cellcolor{lightgray!84}0.84 & \cellcolor{lightgray!78}0.78 & \cellcolor{lightgray!64}0.64 & \cellcolor{lightgray!22}0.22 & \cellcolor{lightgray!86}0.86 & \cellcolor{lightgray!80}0.80 & \cellcolor{lightgray!55}0.55 & \cellcolor{lightgray!19}0.19 & \cellcolor{lightgray!85}0.85 & \cellcolor{lightgray!79}0.79 & \cellcolor{lightgray!59}0.59 & \cellcolor{lightgray!20}0.20 \\
& \qwen &\cellcolor{lightgray!79}0.79 &\cellcolor{lightgray!73}0.73 &\cellcolor{lightgray!82}0.82 &\cellcolor{lightgray!27}0.27 &\cellcolor{lightgray!80}0.80 &\cellcolor{lightgray!74}0.74 &\cellcolor{lightgray!77}0.77 &\cellcolor{lightgray!26}0.26 &\cellcolor{lightgray!80}0.80 &\cellcolor{lightgray!73}0.73 &\cellcolor{lightgray!79}0.79 &\cellcolor{lightgray!27}0.27 \\
\cmidrule{1-14}
\multirow[c]{4}{*}{\rotatebox{90}{Object}} 
& \blip & \cellcolor{lightgray!75}0.75 & \cellcolor{lightgray!67}0.67 & \cellcolor{lightgray!100}1.07 & \cellcolor{lightgray!33}0.33 &\cellcolor{lightgray!79}0.79 & \cellcolor{lightgray!71}0.71 & \cellcolor{lightgray!92}0.92 & \cellcolor{lightgray!28}0.28 & \cellcolor{lightgray!77}0.77 & \cellcolor{lightgray!69}0.69 & \cellcolor{lightgray!100}1.00 & \cellcolor{lightgray!30}0.30 \\
& \deepseek &\cellcolor{lightgray!78}0.78 &\cellcolor{lightgray!71}0.71 &\cellcolor{lightgray!82}0.82 &\cellcolor{lightgray!28}0.29 & \cellcolor{lightgray!78}0.78 &\cellcolor{lightgray!72}0.72 &\cellcolor{lightgray!80}0.80 &\cellcolor{lightgray!30}0.30 &\cellcolor{lightgray!78}0.78 &\cellcolor{lightgray!71}0.71 &\cellcolor{lightgray!81}0.81 &\cellcolor{lightgray!28}0.29 \\
& \lava & \cellcolor{lightgray!84}0.84 & \cellcolor{lightgray!78}0.78 & \cellcolor{lightgray!63}0.63 & \cellcolor{lightgray!21}0.21 & \cellcolor{lightgray!85}0.85 & \cellcolor{lightgray!79}0.79 & \cellcolor{lightgray!59}0.59 & \cellcolor{lightgray!20}0.20 & \cellcolor{lightgray!85}0.85 & \cellcolor{lightgray!78}0.78 & \cellcolor{lightgray!61}0.61 & \cellcolor{lightgray!20}0.20 \\
& \qwen &\cellcolor{lightgray!79}0.79 &\cellcolor{lightgray!72}0.72 &\cellcolor{lightgray!89}0.89 &\cellcolor{lightgray!28}0.28 &\cellcolor{lightgray!79}0.79 &\cellcolor{lightgray!72}0.72 &\cellcolor{lightgray!85}0.85 & \cellcolor{lightgray!28}0.28 &\cellcolor{lightgray!79}0.79 &\cellcolor{lightgray!72}0.72 &\cellcolor{lightgray!87}0.87 &\cellcolor{lightgray!28}0.28 \\
\cmidrule{1-14}
\multirow[c]{4}{*}{\rotatebox{90}{Animal}} 
& \blip & \cellcolor{lightgray!75}0.75 & \cellcolor{lightgray!66}0.66 & \cellcolor{lightgray!100}1.08 & \cellcolor{lightgray!33}0.33 & \cellcolor{lightgray!74}0.74 & \cellcolor{lightgray!65}0.65 & \cellcolor{lightgray!100}1.13 & \cellcolor{lightgray!35}0.35 & \cellcolor{lightgray!74}0.74 & \cellcolor{lightgray!66}0.66 & \cellcolor{lightgray!100}1.10 & \cellcolor{lightgray!34}0.34 \\
& \deepseek &\cellcolor{lightgray!77}0.77 &\cellcolor{lightgray!70}0.70 &\cellcolor{lightgray!84}0.84 &\cellcolor{lightgray!30}0.30 &\cellcolor{lightgray!75}0.75 &\cellcolor{lightgray!68}0.68 &\cellcolor{lightgray!90}0.90 &\cellcolor{lightgray!34}0.34 &\cellcolor{lightgray!76}0.76 &\cellcolor{lightgray!69}0.69 &\cellcolor{lightgray!88}0.88 &\cellcolor{lightgray!32}0.32 \\
& \lava & \cellcolor{lightgray!90}0.90 & \cellcolor{lightgray!86}0.86 & \cellcolor{lightgray!39}0.39 & \cellcolor{lightgray!14}0.14 & \cellcolor{lightgray!86}0.86 & \cellcolor{lightgray!81}0.81 & \cellcolor{lightgray!53}0.53 & \cellcolor{lightgray!20}0.20 & \cellcolor{lightgray!88}0.88 & \cellcolor{lightgray!84}0.84 & \cellcolor{lightgray!46}0.46 & \cellcolor{lightgray!17}0.17 \\
& \qwen &\cellcolor{lightgray!80}0.80 &\cellcolor{lightgray!73}0.73 &\cellcolor{lightgray!80}0.80 &\cellcolor{lightgray!26}0.26 &\cellcolor{lightgray!79}0.79 &\cellcolor{lightgray!72}0.72 &\cellcolor{lightgray!84}0.84 &\cellcolor{lightgray!28}0.28 &\cellcolor{lightgray!80}0.80 &\cellcolor{lightgray!73}0.73 &\cellcolor{lightgray!82}0.82 &\cellcolor{lightgray!27}0.27 \\
\cmidrule{1-14}
\multirow[c]{4}{*}{\rotatebox{90}{Vegetation}}
& \blip & \cellcolor{lightgray!75}0.75 & \cellcolor{lightgray!65}0.65 & \cellcolor{lightgray!100}1.13 & \cellcolor{lightgray!34}0.34 &\cellcolor{lightgray!75}0.75 & \cellcolor{lightgray!66}0.66 & \cellcolor{lightgray!100}1.08 &\cellcolor{lightgray!32}0.32 & \cellcolor{lightgray!75}0.75 & \cellcolor{lightgray!66}0.66 & \cellcolor{lightgray!100}1.11 & \cellcolor{lightgray!33}0.33 \\
& \deepseek &\cellcolor{lightgray!76}0.76 &\cellcolor{lightgray!68}0.68 &\cellcolor{lightgray!92}0.92 &\cellcolor{lightgray!32}0.32 & \cellcolor{lightgray!74}0.74 &\cellcolor{lightgray!67}0.67 &\cellcolor{lightgray!95}0.95 & \cellcolor{lightgray!35}0.35 & \cellcolor{lightgray!75}0.75 &\cellcolor{lightgray!68}0.68 &\cellcolor{lightgray!94}0.94 & \cellcolor{lightgray!33}0.33 \\
& \lava & \cellcolor{lightgray!86}0.86 & \cellcolor{lightgray!80}0.80 & \cellcolor{lightgray!57}0.58 & \cellcolor{lightgray!19}0.19 & \cellcolor{lightgray!86}0.86 & \cellcolor{lightgray!80}0.80 & \cellcolor{lightgray!56}0.56 & \cellcolor{lightgray!19}0.19 & \cellcolor{lightgray!86}0.86 & \cellcolor{lightgray!80}0.80 & \cellcolor{lightgray!56}0.57 & \cellcolor{lightgray!19}0.19 \\
& \qwen &\cellcolor{lightgray!80}0.80 &\cellcolor{lightgray!73}0.73 &\cellcolor{lightgray!84}0.84 &\cellcolor{lightgray!27}0.27 &\cellcolor{lightgray!81}0.81 &\cellcolor{lightgray!75}0.75 &\cellcolor{lightgray!78}0.78 &\cellcolor{lightgray!26}0.26 &\cellcolor{lightgray!80}0.80 &\cellcolor{lightgray!74}0.74 &\cellcolor{lightgray!81}0.81 &\cellcolor{lightgray!26}0.26 \\
\bottomrule
\end{tabular}
\end{table}
For both individual and aggregated datasets, \lava is the most confident and certain model across the four classes. For \trashcan and the aggregated dataset, \blip is generally the least confident and relatively more uncertain, except for MSP on Trash, where \deepseek is slightly less confident; however, the difference is minimal. For \seaclear, the least confident and most uncertain model is mostly \blip: only in one case it ties with \deepseek (PCS for Trash) and in another one with \qwen (DG for Object). In the other cases, differences with the worst model are minimal (e.g., for MSP on Trash, the difference between \deepseek and \blip is only 0.02). Therefore, in general, the differences across models are not very large. For example, on \seaclear, the MSP of \lava is 0.86, while the worst-performing model, \deepseek, has an MSP of 0.77.

\begin{figure}
\centering 
\begin{subfigure}[b]{1\columnwidth}
\centering
\includegraphics[width=0.5\linewidth]{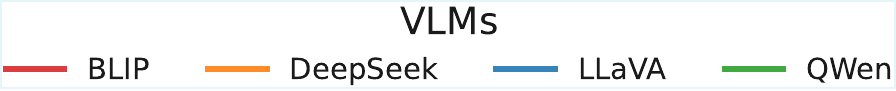}
\end{subfigure}


\begin{subfigure}[b]{0.3\columnwidth}
\centering
\includegraphics[width=1\linewidth]{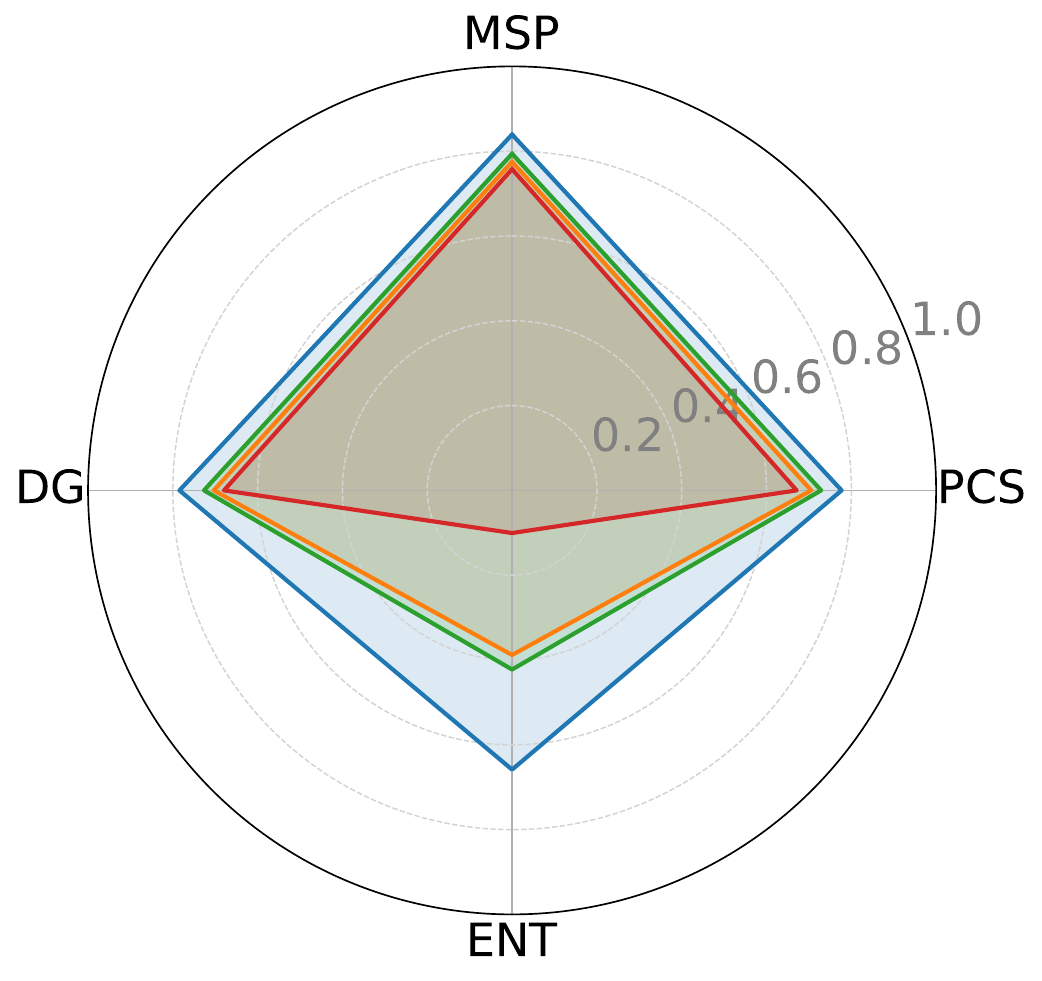}
\caption{TrashCan1.0 -- Trash}
\label{fig:rq2_perclass_trash_trashcan}
\end{subfigure}
\hfill%
\begin{subfigure}[b]{0.3\columnwidth}
\centering
\includegraphics[width=1\linewidth]{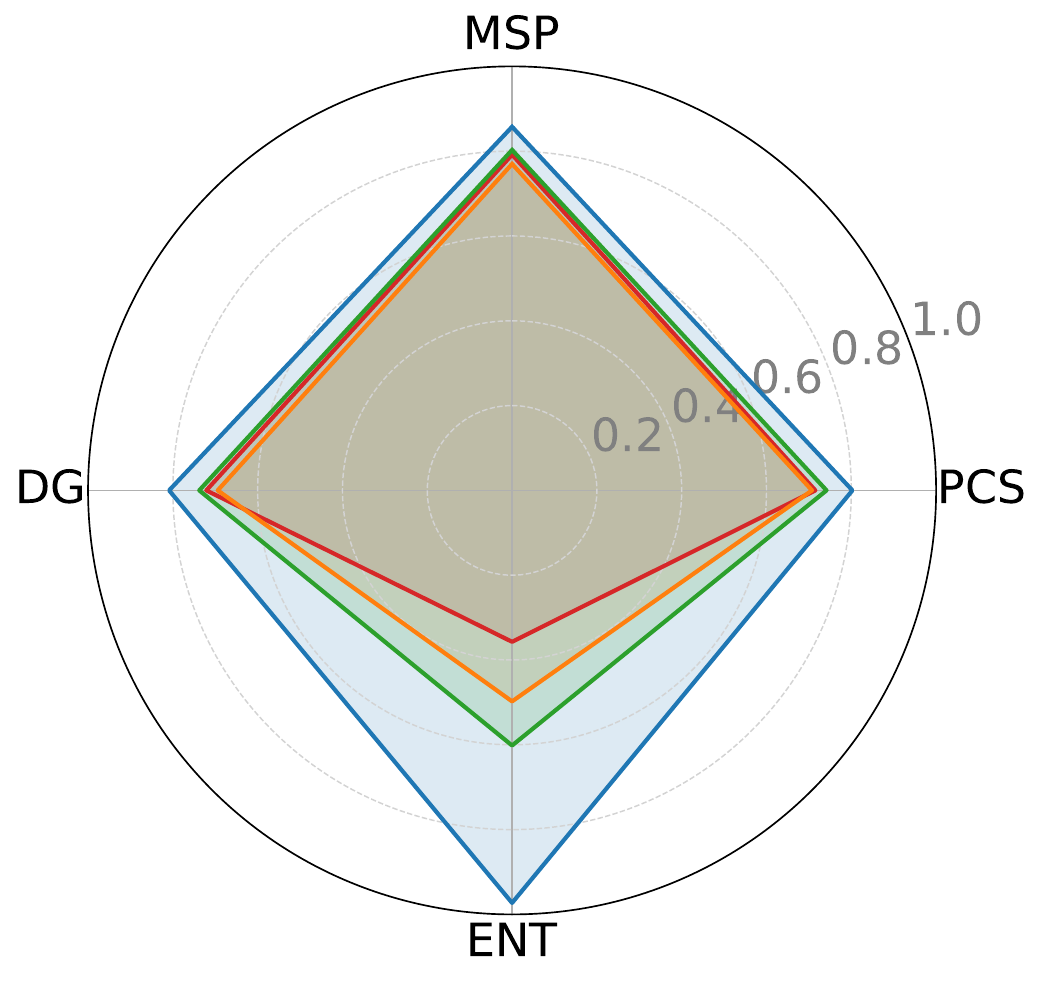}
\caption{SeaClear -- \quad Trash}
\label{fig:rq2_perclass_trash_seaclear}
\end{subfigure}
\hfill%
\begin{subfigure}[b]{0.3\columnwidth}
\centering
\includegraphics[width=1\linewidth]{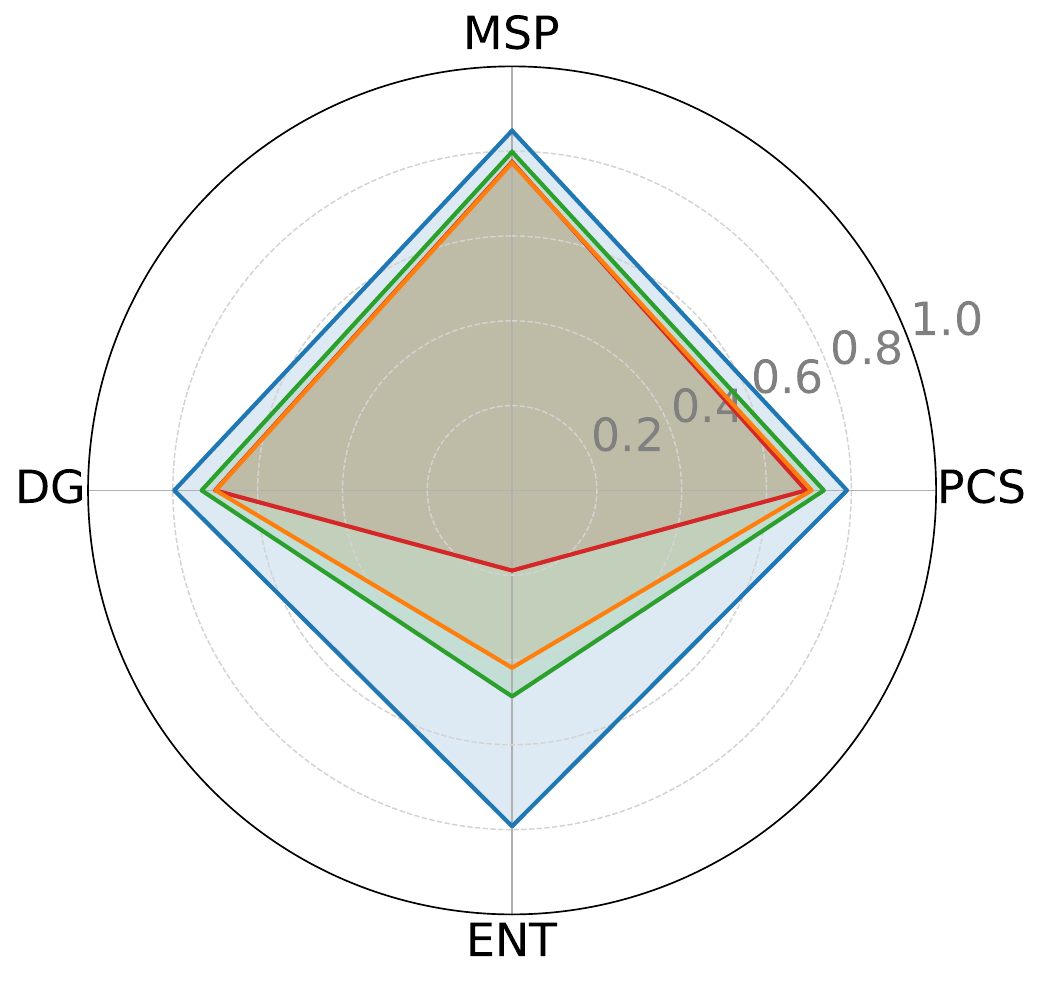}
\caption{Aggregated -- Trash}
\label{fig:rq2_perclass_trash_aggregate}
\end{subfigure}

\begin{subfigure}[b]{0.3\columnwidth}
\centering
\includegraphics[width=1\linewidth]{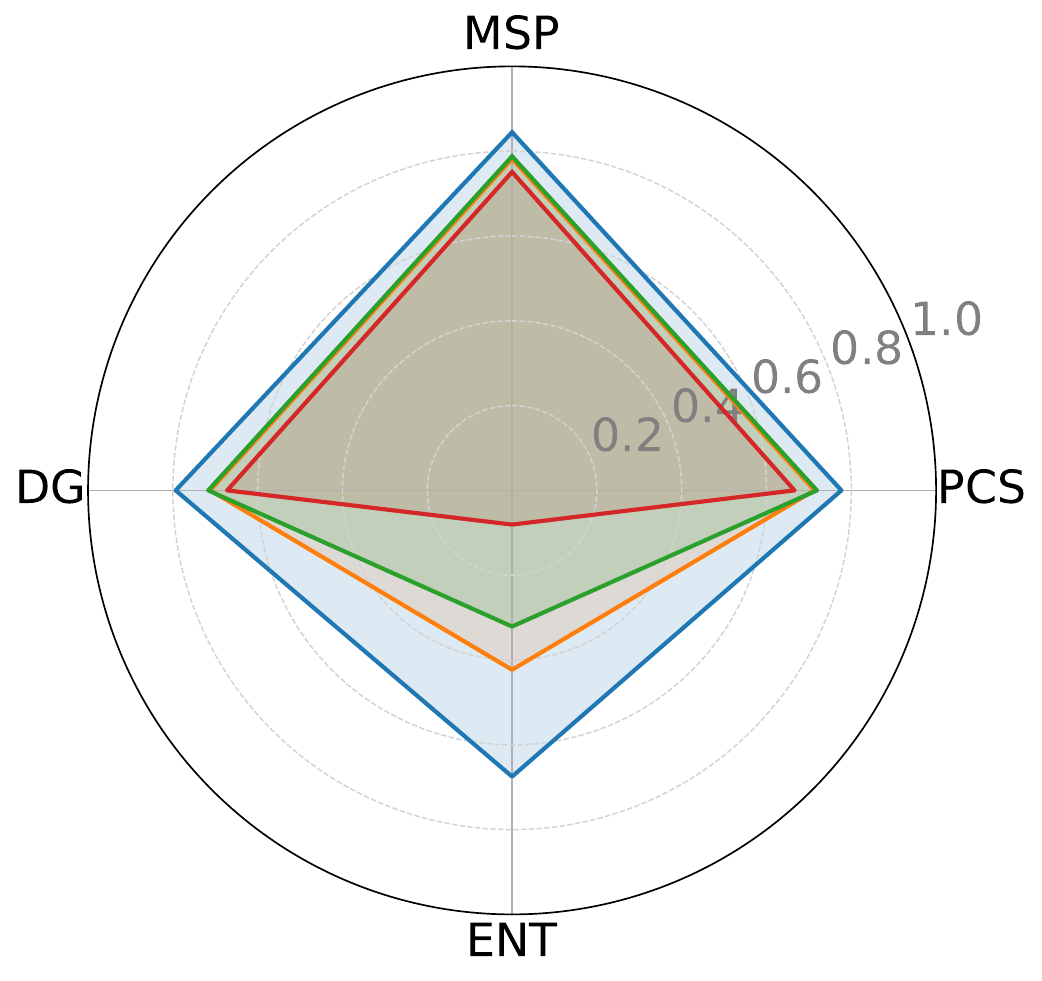}
\caption{TrashCan1.0 -- Object}
\label{fig:rq2_perclass_object_trashcan}
\end{subfigure}
\hfill%
\begin{subfigure}[b]{0.3\columnwidth}
\centering
\includegraphics[width=1\linewidth]{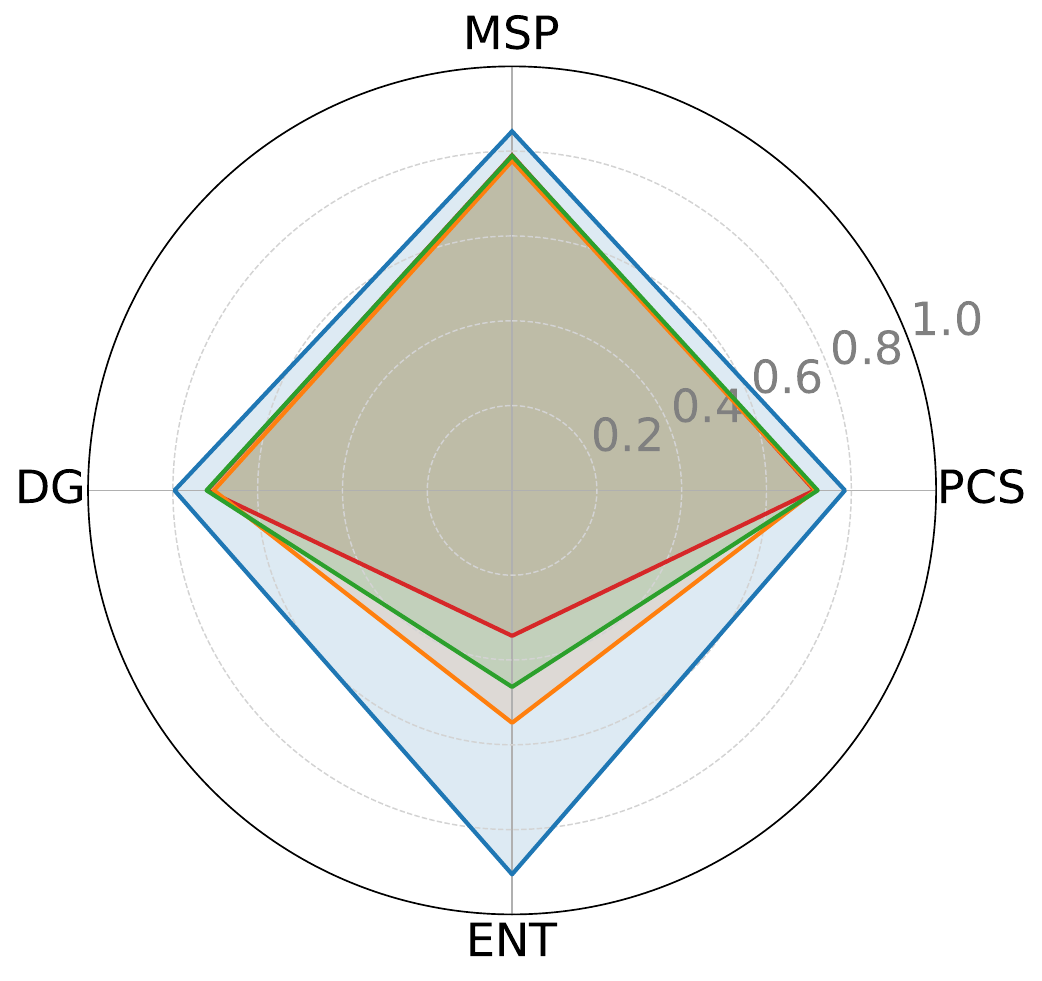}
\caption{SeaClear -- \quad Object}
\label{fig:rq2_perclass_object_seaclear}
\end{subfigure}
\hfill%
\begin{subfigure}[b]{0.3\columnwidth}
\centering
\includegraphics[width=1\linewidth]{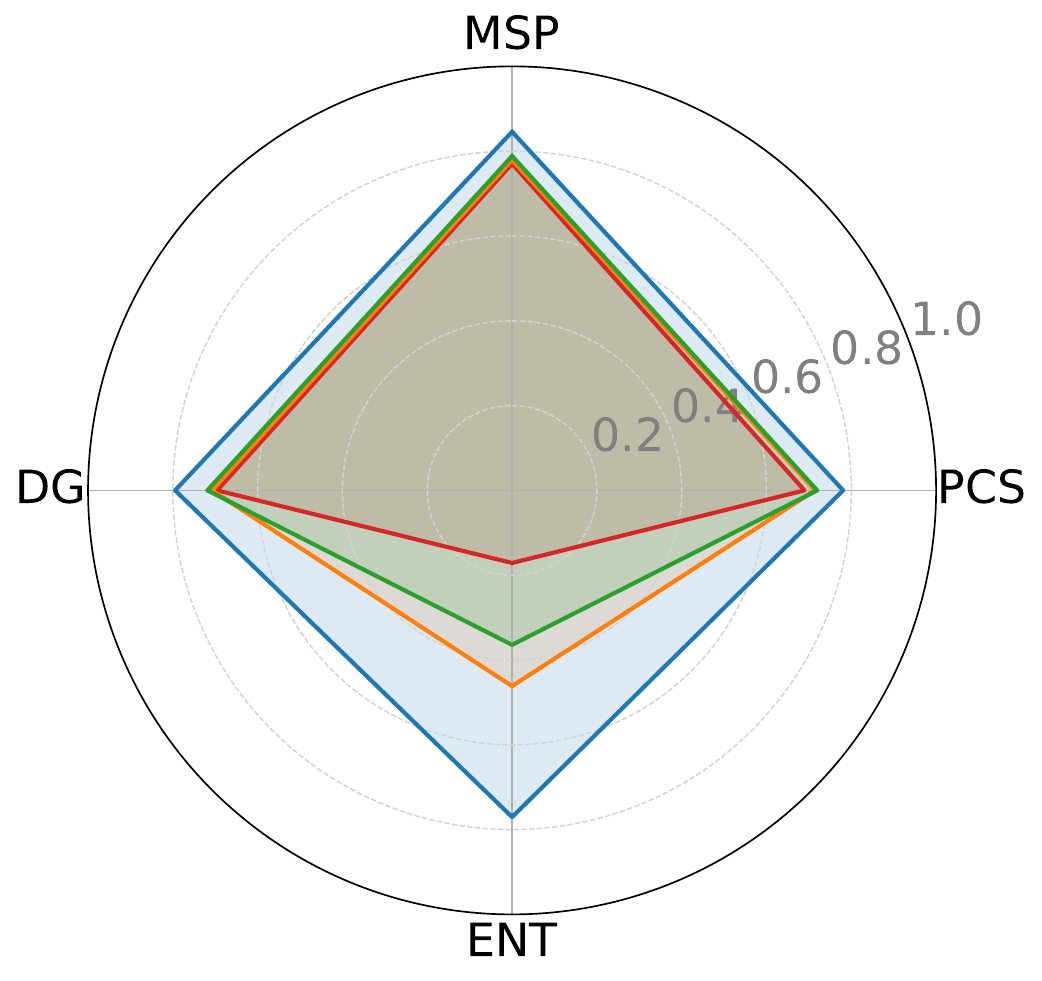}
\caption{Aggregated -- Object}
\label{fig:rq2_perclass_object_aggregate}
\end{subfigure}
\begin{subfigure}[b]{0.3\columnwidth}
\centering
\includegraphics[width=1\linewidth]{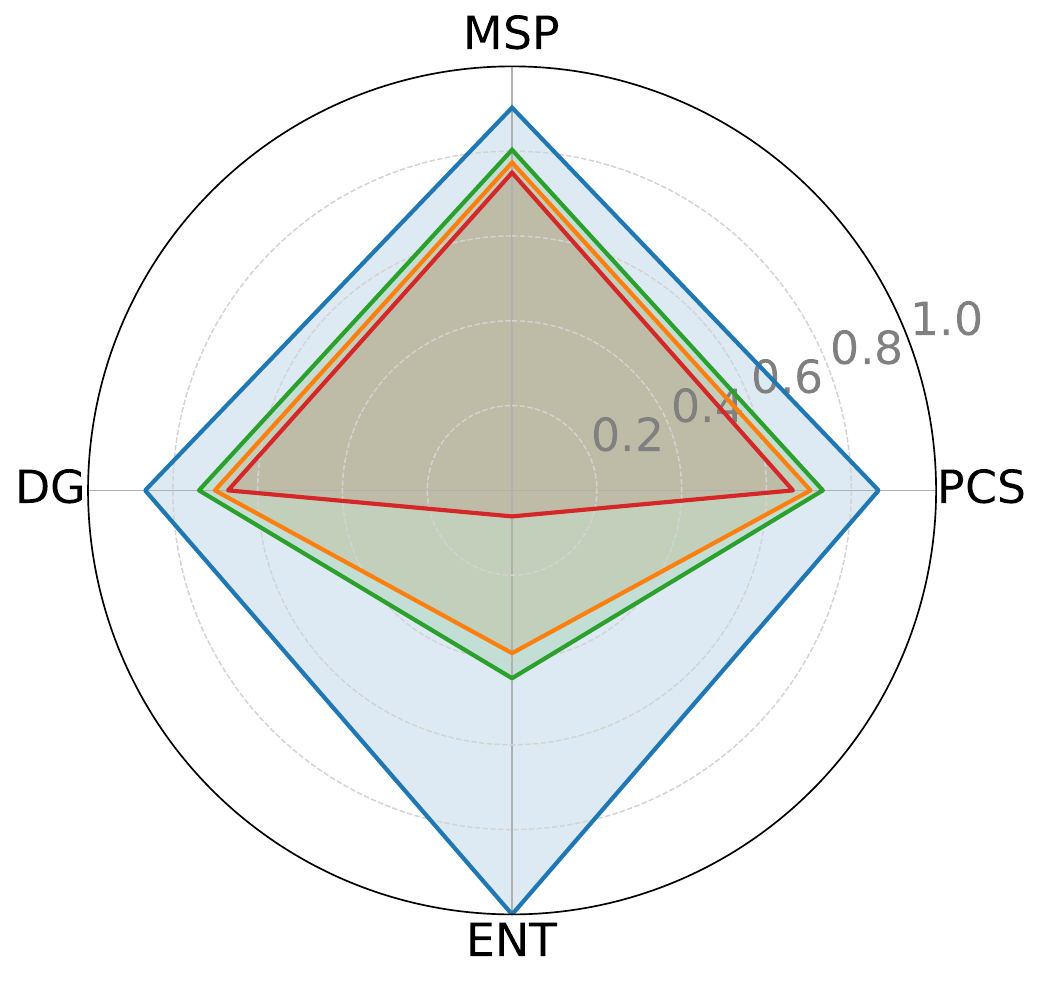}
\caption{TrashCan1.0 -- Animal}
\label{fig:rq2_perclass_animal_trashcan}
\end{subfigure}
\hfill%
\begin{subfigure}[b]{0.3\columnwidth}
\centering
\includegraphics[width=1\linewidth]{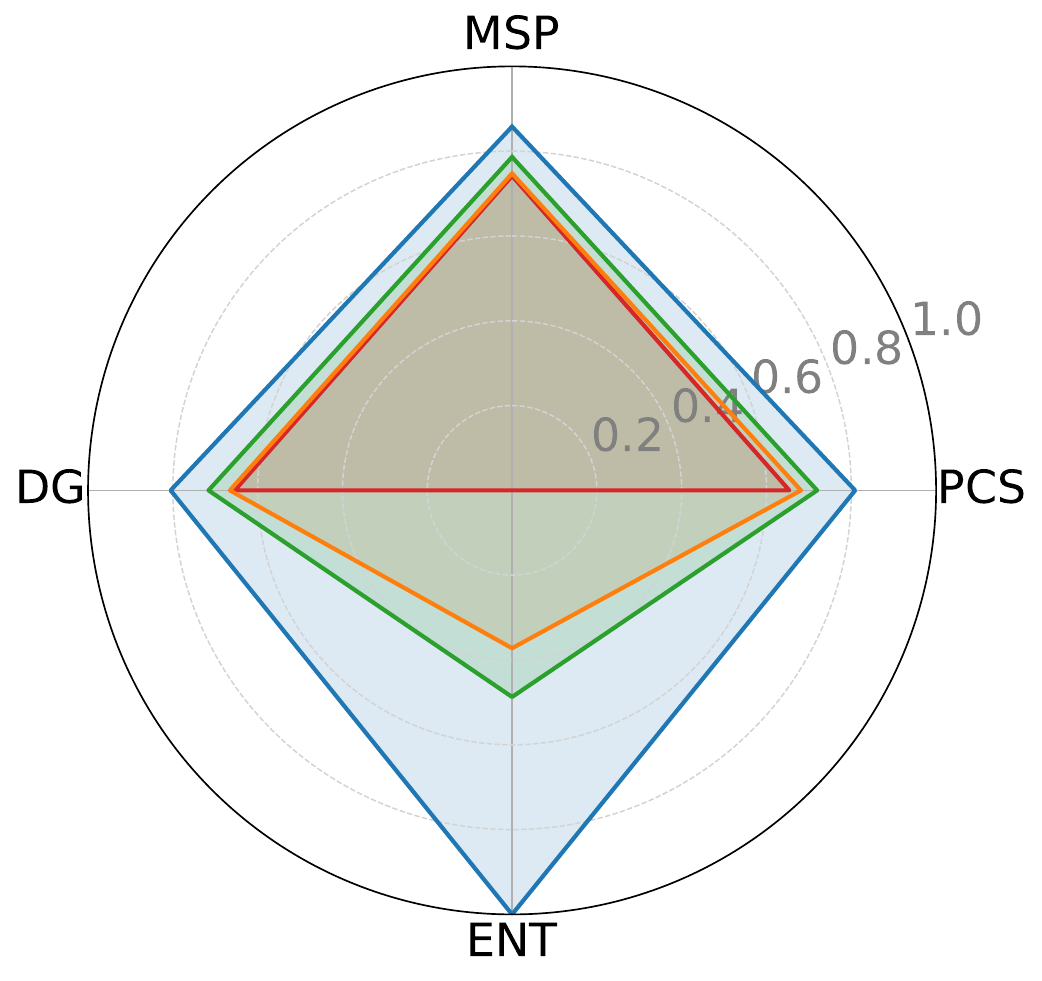}
\caption{SeaClear -- Animal}
\label{fig:rq2_perclass_animal_seaclear}
\end{subfigure}
\hfill%
\begin{subfigure}[b]{0.3\columnwidth}
\centering
\includegraphics[width=1\linewidth]{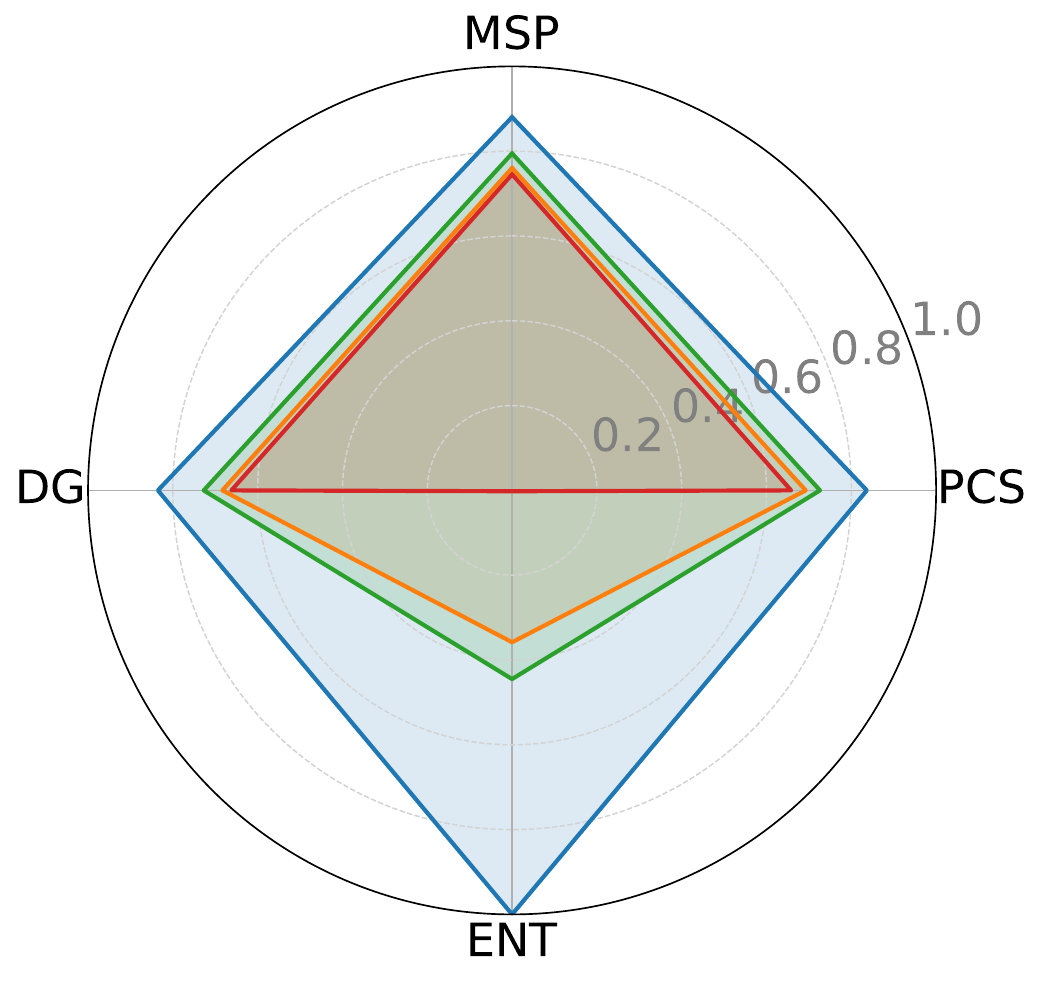}
\caption{Aggregated -- Animal}
\label{fig:rq2_perclass_animal_aggregate}
\end{subfigure}
\begin{subfigure}[b]{0.3\columnwidth}
\centering
\includegraphics[width=1\linewidth]{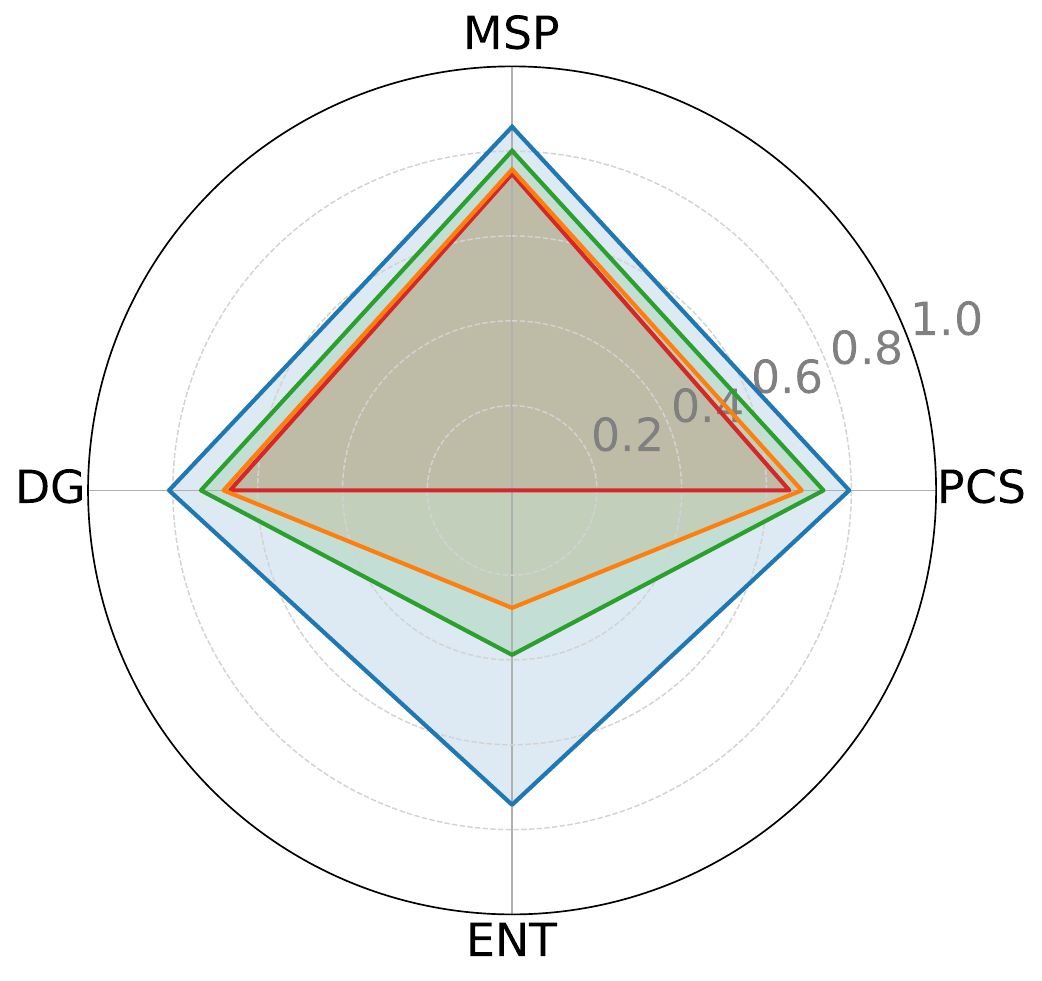}
\caption{TrashCan1.0 -- Vegetation}
\label{fig:rq2_perclass_veget_trashcan}
\end{subfigure}
\hfill%
\begin{subfigure}[b]{0.3\columnwidth}
\centering
\includegraphics[width=1\linewidth]{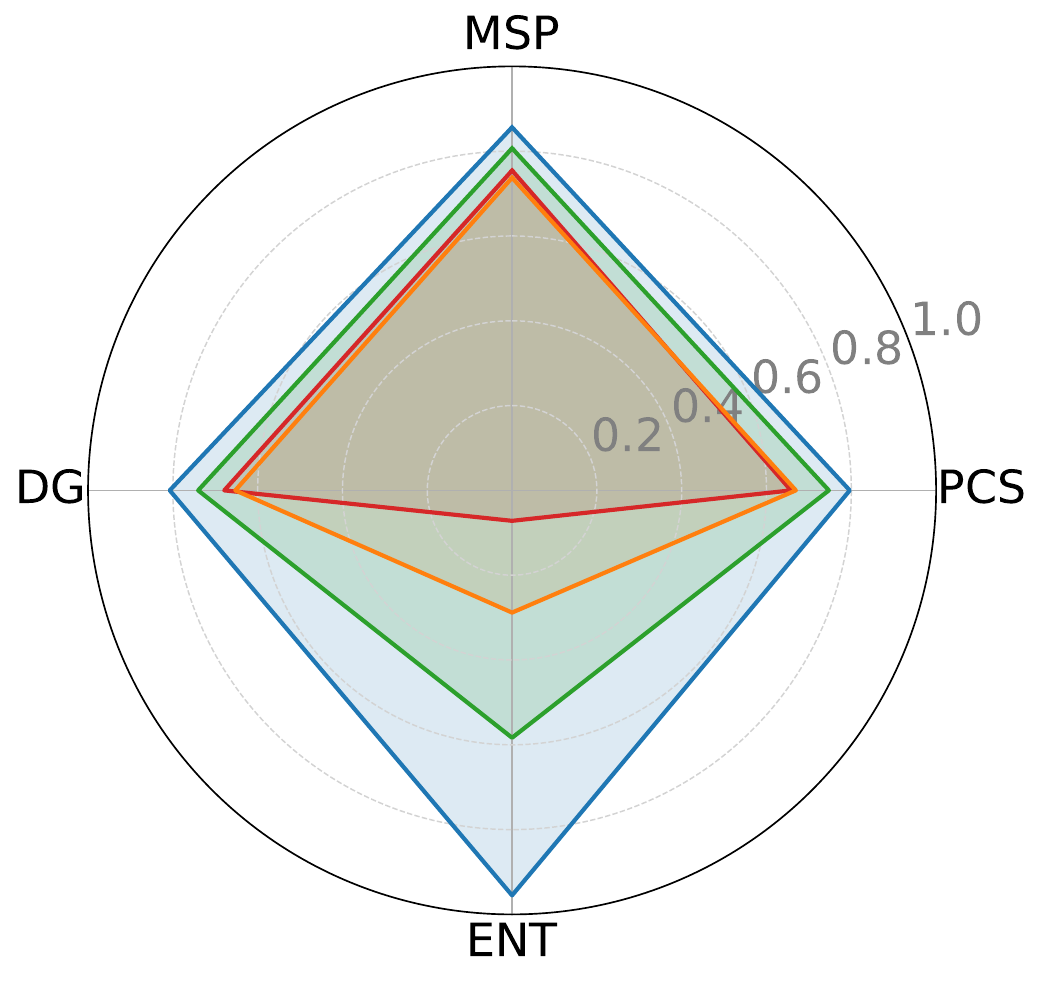}
\caption{SeaClear -- Vegetation}
\label{fig:rq2_perclass_veget_seaclear}
\end{subfigure}
\hfill%
\begin{subfigure}[b]{0.3\columnwidth}
\centering
\includegraphics[width=1\linewidth]{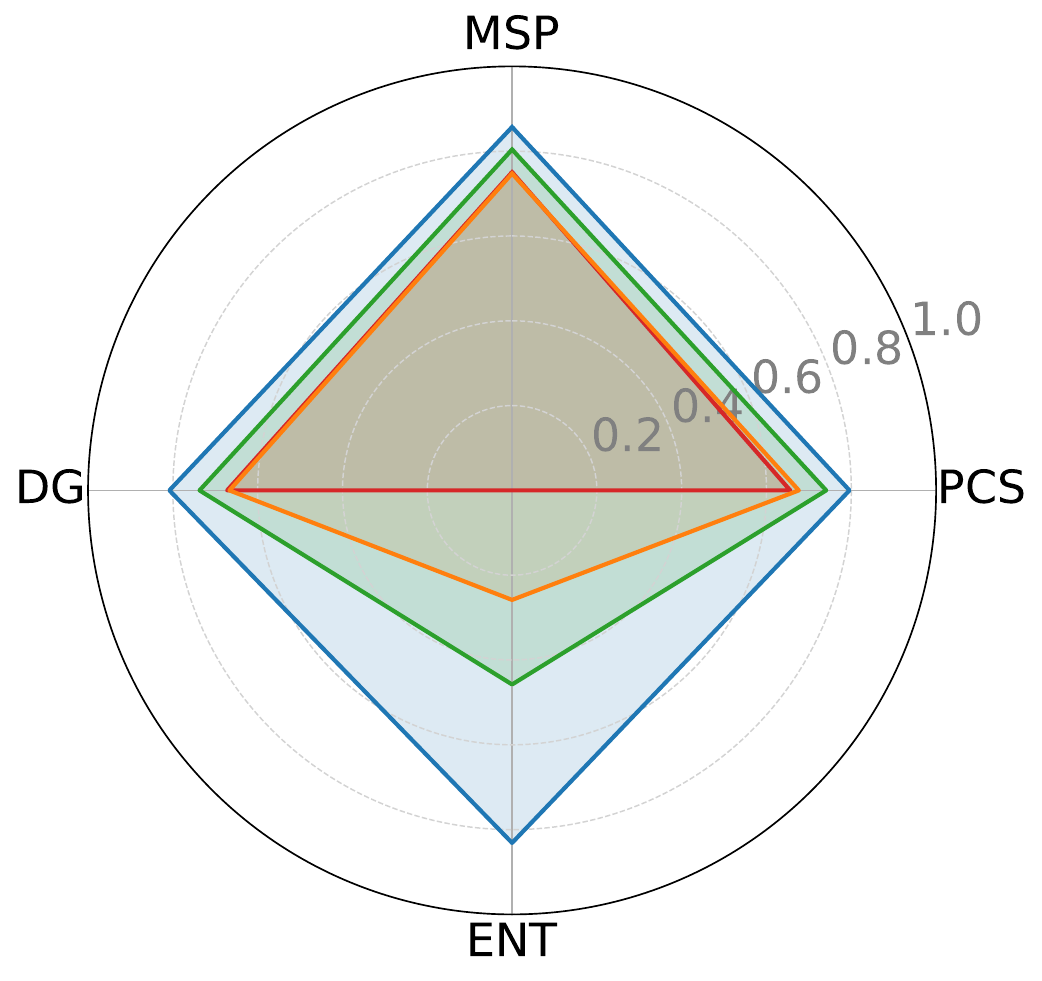}
\caption{Aggregated -- Vegetation}
\label{fig:rq2_perclass_veget_aggregate}
\end{subfigure}
\caption{RQ2 -- Uncertainty of the VLMs across datasets for each class. Values are normalized and scaled to the same range: higher values indicate higher confidence (MSP and PCS) and lower uncertainty (DG and ENT).}
\label{fig:rq2_perclass}
\end{figure}

We also present the per-class uncertainty quantification results in \Cref{fig:rq2_perclass}
for both individual and aggregated datasets across the four classes.
%
Note that values for DG and ENT are used in original form without normalization and to be interpreted unlike confidence metrics, i.e., lower values indicate lower uncertainty, with lower uncertainty being desirable. For the Trash, Animal, and Vegetation classes shown in 
\Cref{fig:rq2_perclass},
\lava is the most confident and certain model, followed by \qwen, across both individual and aggregated datasets. For Object, although \lava remains the most confident and certain model, \deepseek and \qwen are the second most confident and certain for \trashcan and the aggregated dataset, while for \seaclear, \qwen, \lava, and \deepseek show similar results. Overall, \blip remains the least confident and more uncertain, consistent with the observations in Table~\ref{tab:RQ2_per_class_uq_summary}. However, one can observe that the differences between models in terms of MSP, PCS, and DG are relatively small. The only exception is ENT, where \lava shows considerably larger differences compared to the other models.

\finding{\textbf{RQ2:} \lava is consistently the most confident and certain model across individual and aggregated datasets, but it is poorly calibrated, indicating overconfidence. In contrast, \blip is generally the least confident and more uncertain, though the differences from \lava are minor, except in terms of entropy—yet it is the best-calibrated model, making it cautious and reliable.}

\subsection{RQ3 -- Relationship Between Performance and Uncertainty}
\begin{figure}
\centering 
\includegraphics[width=\linewidth]{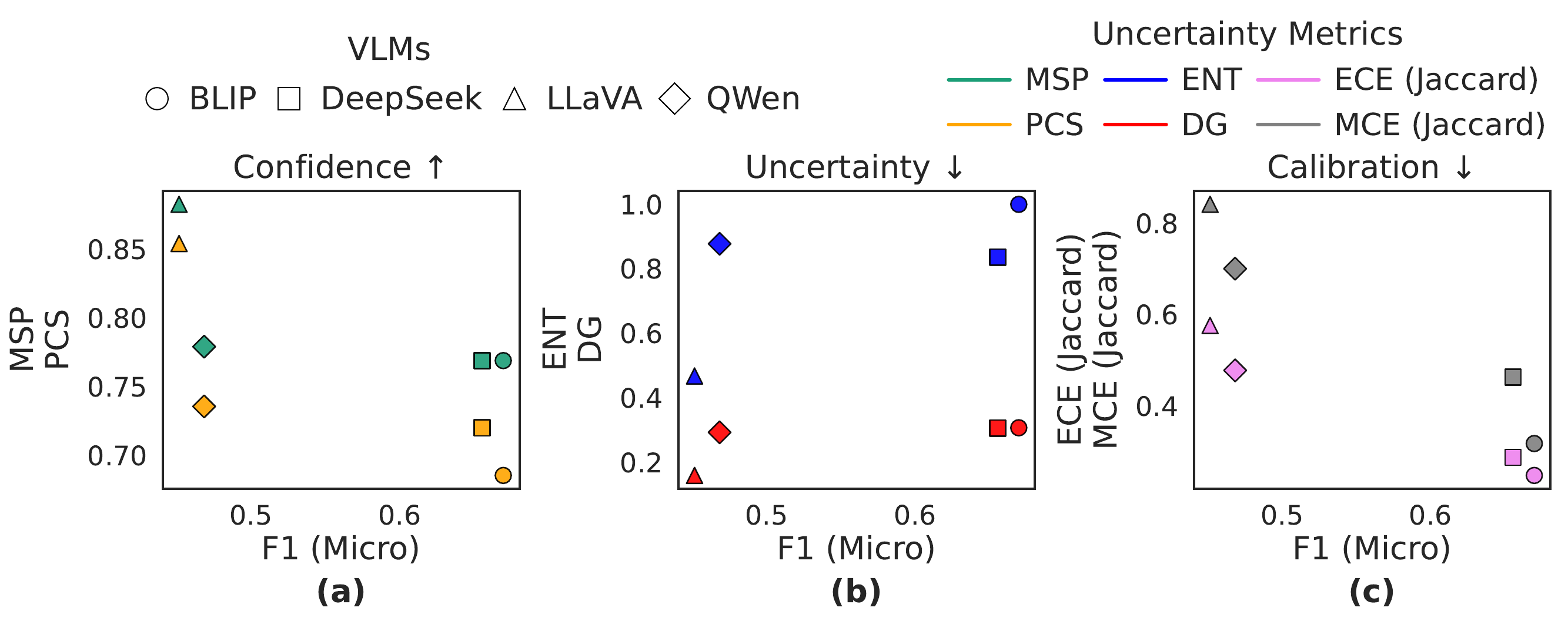}
\caption{RQ3 -- Comparison of F1 (Micro) vs Uncertainty Metrics for four VLMs across Aggregated Dataset}
\label{fig:rq3_f1micro}
\end{figure}

\begin{figure}[!tb]
\centering 
\includegraphics[width=\linewidth]{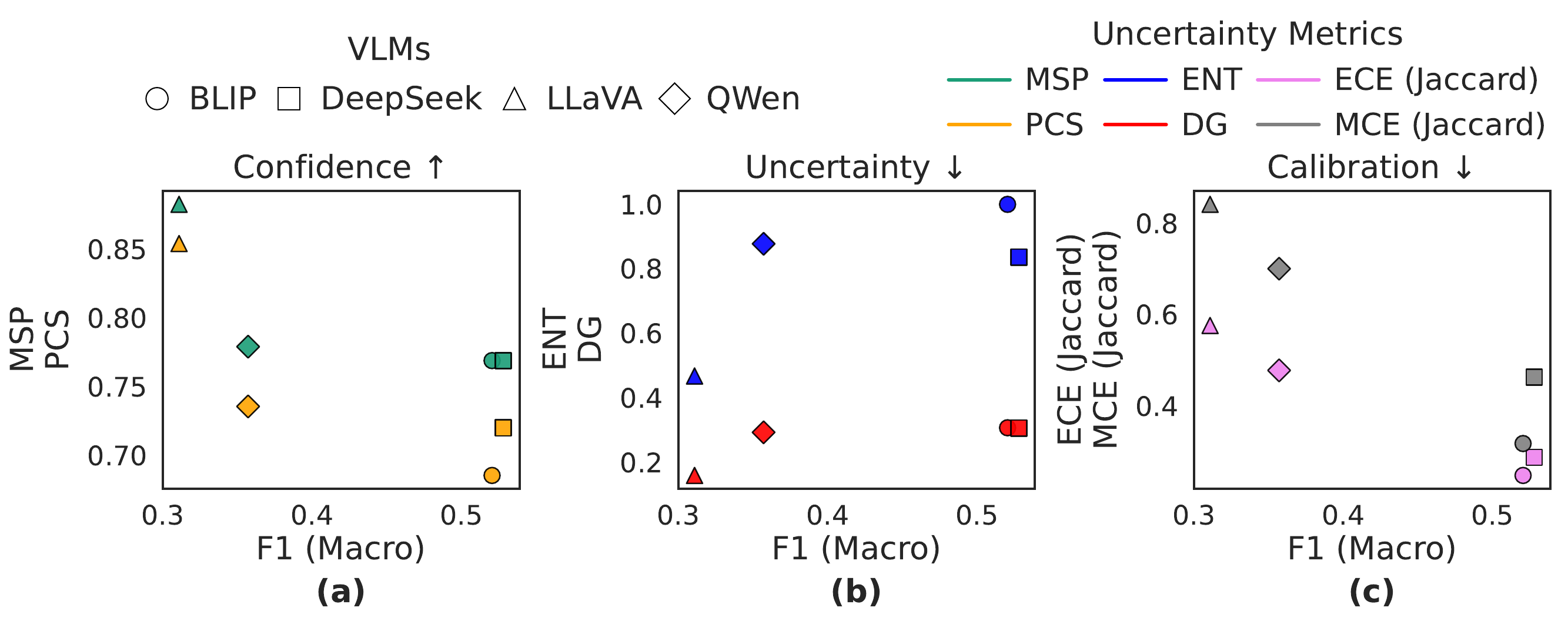}
\caption{RQ3 -- Comparison of F1 (Macro) vs Uncertainty Metrics for four VLMs across Aggregated Dataset}
\label{fig:rq3_f1macro}
\end{figure}

\begin{figure}[!tb]
\centering 
\includegraphics[width=\linewidth]{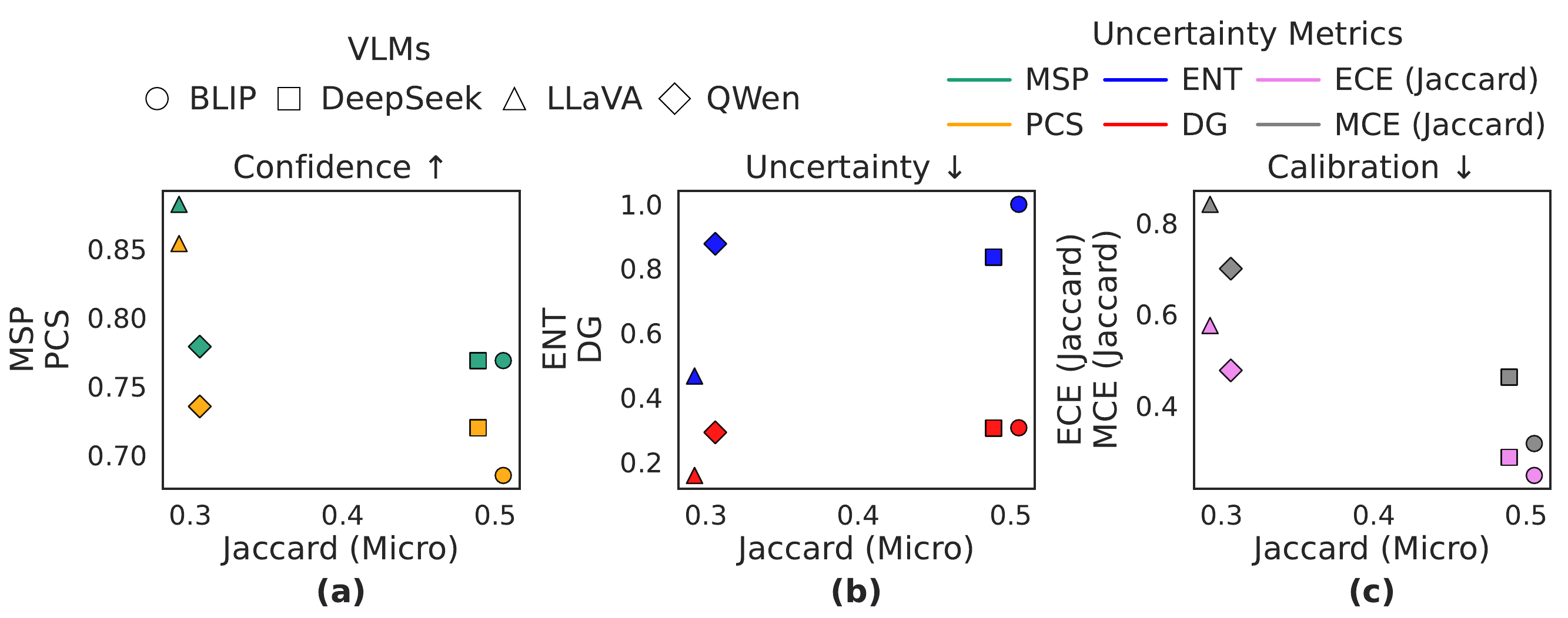}
\caption{RQ3 -- Comparison of Jaccard Micro vs Uncertainty Metrics for four VLMs across Aggregated Dataset}
\label{fig:rq3_jmicro}
\end{figure}

\begin{figure}[!tb]
\centering
\includegraphics[width=\linewidth]{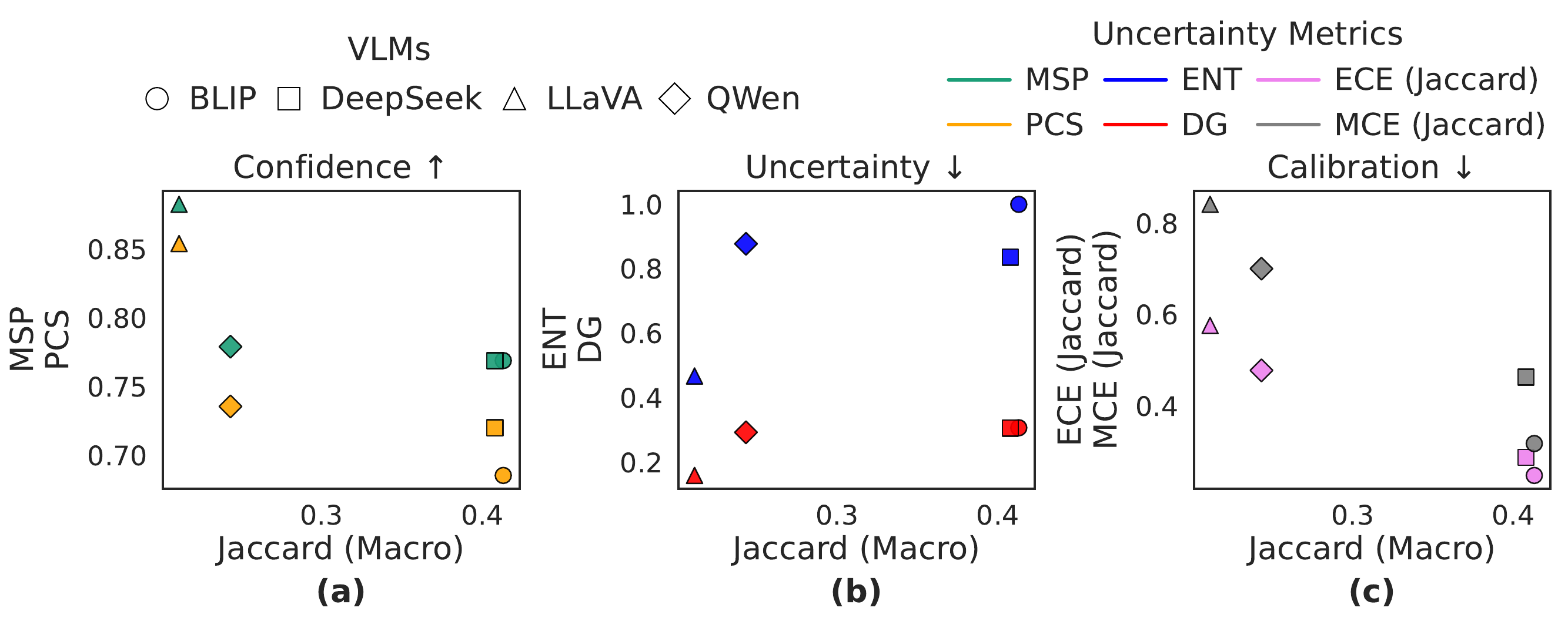}
\caption{RQ3 -- Comparison of Jaccard Macro vs Uncertainty Metrics for four VLMs across Aggregated Dataset}
\label{fig:rq3_jmacro}
\end{figure}

We study how VLM performance relates to uncertainty. \Cref{fig:rq3_f1micro}~(a) shows the relationships between F1 (Micro) and the confidence metrics (MSP and PCS).
%
The results indicate that both \blip and \deepseek achieve the highest performance, as also shown in RQ1, but they exhibit slightly lower confidence in MSP and PCS than the most confident VLM, \lava. However, these differences are not substantial. \lava remains the most confident model but shows lower performance, whereas \qwen has confidence levels similar to \blip and \deepseek but lower performance. The results for F1 (Macro), Jaccard (Micro), and Jaccard (Macro) show similar trends in Figs.~\ref{fig:rq3_f1macro}, \ref{fig:rq3_jmicro}, and \ref{fig:rq3_jmacro} respectively. Overall, these findings suggest that \blip and \deepseek are the most suitable models for our context, as they achieve the best performance among the models while maintaining generally high confidence.

Fig.~\ref{fig:rq3_f1micro}~(b) shows the results of F1 (Micro) with the uncertainty metrics (ENT and DG). With respect to performance, \blip and \deepseek are the best, while \qwen and \lava have lower performance. Regarding uncertainty, \lava exhibits the lowest uncertainty in terms of both ENT and DG, while \blip shows the highest uncertainty in terms of ENT and has DG values similar to \deepseek and \qwen, but higher than \lava. These results indicate that the highest-performing models tend to have relatively higher uncertainty in their decisions. In terms of ENT, the differences between these models and the one with the lowest uncertainty are relatively large, whereas for DG, the differences are smaller. Similar results are observed for F1 (Macro), Jaccard (Micro), and Jaccard (Macro) shown in \Cref{fig:rq3_f1macro}~(b), \Cref{fig:rq3_jmicro}~(b), and \Cref{fig:rq3_jmacro}~(b).
%
%
%

\Cref{fig:rq3_f1micro}~(c) shows the results of F1 (Micro) with the calibration metrics (ECE and MCE). Performance-wise, as before, \blip and \deepseek are the best, whereas \qwen and \lava are the worst-performing. In terms of both ECE and MCE, \blip and \deepseek are better calibrated than \qwen and \lava, which exhibit the highest errors. Similar results hold for F1 (Macro), Jaccard (Micro), and Jaccard (Macro), as shown in \Cref{fig:rq3_f1macro}~(c), \Cref{fig:rq3_jmicro}~(c), and \Cref{fig:rq3_jmacro}~(c), respectively. Overall, these results suggest that the best-performing models in our study are also the best-calibrated, whereas the worst-performing models are the least calibrated.

Based on the results, we can conclude that \blip and \deepseek are the best-performing and best-calibrated models, while exhibiting relatively low confidence and high uncertainty, though their confidence and certainty are still comparable to those of the most confident and certain model (i.e., \lava), which is also the worst-performing.

\finding{\textbf{RQ3:} High confidence or low uncertainty in a VLM does not necessarily indicate high performance; instead, high performance is more related to good calibration. Therefore, when selecting VLMs for autonomous underwater robots for trash collection tasks, one should prioritize VLM performance and calibration over confidence and certainty.}

\section{Threats to Validity} \label{sec: threats}
\textbf{Internal Validity.} All experiments were run on Ubuntu 22.04 in Python virtual environments with identical dependencies, consistent GPU allocation, and uniform preprocessing to avoid any bias. Uncertainty metrics were computed directly from logits and token probabilities, with standardized JSON outputs to ensure that uncertainty calculation is not affected by post-hoc manipulation. We averaged token-level metrics across generated tokens per image and aggregated per class and dataset to avoid the effects of randomness. Across \blip, \deepseek, \qwen, and \lava, we used identical prompts, inference settings, and a fixed maximum token length of 80 to minimize configuration-induced differences.

\noindent\textbf{External validity.} Results are based on only two underwater benchmarks, i.e., \trashcan, \seaclear, and their aggregated variant, covering objects, trash, animals, and vegetation. These two datasets are the most comprehensive publicly available. Nonetheless, more extensive studies using additional datasets are needed to generalize the results further. All models were used in zero-shot settings without fine-tuning on deep underwater imagery, which may limit domain-specific generalization; however, this configuration was consistent across all models. To mitigate dataset-specific biases, we macro-averaged results across the datasets.

\noindent\textbf{Construct Validity.} To answer RQ1---RQ3, we used established metrics from performance measurement, uncertainty quantification, and VLM literature. Because MSP and PCS primarily reflect confidence rather than epistemic uncertainty, we complemented them with Entropy, Deep Gini, and calibration metrics (ECE, MCE) to capture dispersion and reliability. All metrics were computed per image and per class to enhance interpretability and limit aggregation bias.

\noindent\textbf{Conclusion Validity.} 
We averaged token-level uncertainty and calibration metrics across tokens per image and then across the dataset to reduce token noise. Minor stochastic variation from numerical precision or token sampling was minimized using deterministic inference (temperature = 0, do\_sample = False). Since VLM inference on the same prompt can still be stochastic, we executed each prompt almost five times and reported the resulting variability into our results (see Sect.~\ref{sec:results}).

\section{Discussion and Lessons Learned}

\paragraph*{\bf VLM's Performance must be Interpreted in a Specific Application Context}

Our empirical evaluation showed that VLMs can support AUR tasks. However, their effectiveness depends on the target application and the object categories. Although our best-performing model, \blip, achieved relatively good overall performance (F1(Micro) = 0.67 on the aggregated dataset), this level of performance indicates that current VLMs are not yet suitable for direct, standalone deployment in industrial AUR software for detecting any object type. Moreover, all evaluated VLMs performed poorly on animal and vegetation classes, highlighting important limitations of VLMs in domain coverage. These findings suggest that industrial deployment of VLMs should be restricted to scoped tasks, such as trash and object identification, where acceptable performance levels were observed. In particular, \blip achieved an F1 score of 0.76 for trash and 0.87 for object classes on the aggregated dataset, indicating that such use cases are more viable in practice. However, in our empirical evaluation, we used the VLMs as-is, and lightweight fine-tuning is likely to improve their performance. \looseness=-1

\paragraph*{\bf Varied Performance on Different Object Classes is a Limitation for Industrial Use}

We observed a large performance gap across object classes, suggesting that VLMs without fine-tuning do not necessarily cover all classes relevant to underwater tasks uniformly. On the aggregated dataset, \blip achieved an F1 (Micro) score of 0.67 overall, with better performance on trash (F1 = 0.76) and object classes (F1 = 0.87), but poor performance on animal and vegetation classes (F1 = 0.28 and 0.17). These results indicate that, in an industrial context, evaluating VLM quality for each object class individually, rather than relying solely on aggregate performance across all classes, is critical to ensure dependable deployment in AUR applications.

\paragraph*{\bf VLMs as Supporting Software Components for AUR Software}

Based on our results, we suggest that VLMs are currently better suited as supporting components of AUR software rather than as the primary perception software. For example, a potential industrial scenario is to use VLMs when deep learning models or other perception implementations exhibit high uncertainty or fail to classify an object. This hybrid approach aligns better with industrial practices and allows VLMs to add value while limiting risk. For on-board deployment, smaller VLM versions are preferred due to limited processing power. This consideration motivated our choice of VLMs with a relatively small number of parameters for the evaluation. \looseness=-1

\paragraph*{\bf VLMs in Digital Twins of AURs}
An industrial use case is the integration of VLMs into the digital twins of AURs, which operate in real time alongside the physical AUR to support perception and classification tasks. In this use case, larger VLMs can be employed, as digital twins can be deployed on powerful computational resources, either locally or in the cloud. For example, if an AUR cannot identify an object underwater, it can request its digital twin to perform the identification and send the results back before taking any physical action on the object (e.g., removing trash). Similarly, digital twins with VLMs can be used to test and validate AUR perception modules prior to on-board deployment, ensuring that well-tested and reliable perception software is installed.

\paragraph*{\bf Uncertainty and Calibration are Important for VLM Deployment in AUR Software}
Our results show that uncertainty and calibration are critical for assessing the performance of VLMs in underwater trash detection. For example, \lava and \qwen generally exhibited high confidence despite low performance and poor calibration, while \blip demonstrated better calibration along with good performance and comparable confidence. These findings indicate that evaluation of VLMs should consider performance alongside calibration and uncertainty to ensure reliable deployment in AUR software.

\paragraph*{\bf Token-Level Uncertainty to Support Testing AUR Software}
Our results showed that token-level uncertainty metrics correlate with performance and act as indicators of potential failures (e.g., mislabeled predictions). From a testing perspective, these metrics can serve as the basis for devising uncertainty-aware testing strategies to find failures during offline testing of VLM-based perception modules for AUR software. Also, they could be used in real time to detect failures and suggest corrective actions, which is particularly valuable in safety-critical domains such as underwater operations. \looseness=-1


\paragraph*{\bf Ensuring Safety and Trustworthiness of VLMs in Maritime is Key}
A key lesson from our empirical evaluation is that safety and trustworthiness must be explicitly considered when integrating VLMs into maritime applications where failures may have severe safety, environmental, and economic consequences. Maritime operations usually involve complex, safety-critical interactions among human operators, software/hardware systems, and dynamic operating environments. Consequently, VLM outputs may directly influence decision-making related to navigation, situational awareness, and emergency response. Our study highlights that, to support industrial certification and assurance, VLM-enabled systems must be reliable and robust under uncertain inputs and environmental variability, and our empirical evaluation provides evidence of VLM performance under these conditions.

\paragraph*{\bf Regulatory Compliance in Maritime should Align with Technical Evolution}
The maritime industry operates within a highly regulated environment governed by the International Maritime Organization (IMO) and classification rules established by classification societies. Maritime companies are therefore required to comply with a strict regulatory framework to ensure operational safety and security. Our empirical evaluation provides initial evidence that may be relevant for demonstrating regulatory compliance when integrating  VLMs as components into maritime software, which often evolves faster than existing regulations, creating compliance challenges for industry stakeholders. This highlights the importance of robust, well-tested VLMs to ensure regulatory compliance when employing VLMs in practice.

\section{Related Work}\label{sec: related_work}


AURs are becoming increasingly important for several marine tasks, e.g., underwater environmental monitoring, inspection, and trash detection~\cite{drever2018double}. Earlier AURs mainly relied on humans remotely operating them for marine tasks with limited autonomy due to the complexity of underwater conditions, e.g., poor lighting, turbidity, visual noise, and fluid dynamics~\cite{yuan2017aekf, goncalves2022operational, saoud2023mars}. Such conditions severely degrade AUR's perception and manipulation~\cite{walia2025deep}. Recent advances in computer vision and deep learning have enhanced perception and decision-making, paving the way for semi- or fully AURs.

Progress has been achieved in deep learning for marine tasks, including pollution detection~\cite{kshirsagar2021ocean}, trash detection~\cite{xue2021efficient, bajaj2021sea}, submerged debris detection~\cite{xue2021deepsea}, and localization of trash~\cite{sanchez2023experimental}. 
These efforts demonstrate the efficacy of deep learning in identifying underwater trash, with innovations such as the {\it YOLOTrashCan} network emerging as a dominant tool for marine debris identification~\cite{zhou2022yolotrashcan}. While deep learning excels at classification, it struggles to recognize objects outside its training set. In contrast, VLMs can identify unseen objects through their strong generalization ability. As a first step toward their use in AURs, we empirically evaluate their performance in underwater object detection.


Recent developments in open-source VLMs have revolutionized multimodal understanding by combining visual perception and natural language reasoning. General-purpose VLMs such as CLIP~\cite{Radford2021CLIP}, BLIP-2\cite{Li2022BLIP}, LLaVa\cite{liu2024Lava}, and DeepSeek-VL\cite{Liu2024DeepSeekV3} show strong performance and generalization capabilities, including feature extraction, captioning, and visual reasoning in zero-shot settings. Multi-modality VLMs pre-trained on text-image pairs can perform object classification, scene captioning, and trash detection effectively without additional training cost, offering a good alternative to supervised CNN-based detection networks~\cite{Kuo2022FVLM}. However, their ability to detect objects in underwater images has not been studied in the literature, which this paper addresses.

Domain‑adapted models such as MarineGPT\cite{Zheng2023MarineGPT} and AquaticCLIP fine‑tune open‑source VLMs on large corpora of underwater image–text pairs, achieving strong performance on marine organism classification, captioning, and description tasks. MarineInst further scales to \textasciitilde20M instances for marine images, improving segmentation, captioning, and semantic descriptions~\cite{Zheng2024MarineInst}. However, these models are primarily trained for marine organism categories rather than object/trash detection. Moreover, they do not study the uncertainty and miscalibration of VLMs. In contrast, we evaluate four VLMs for multi-label classification, assessing their zero-shot performance and uncertainty estimates across trash, objects, animals, and vegetation classes relevant to AURs.

Quantification of VLM's uncertainty is essential for making trustworthy decisions. VLMs are often overconfident under noise or domain shift, reflecting aleatoric (data) and epistemic (model) uncertainties~\cite{Abdar2021UQReview}. Recent benchmarks show weak correlations between uncertainty and accuracy in generative and multimodal models~\cite{Shorinwa2025UQLLMSurvey}. Classical uncertainty quantification methods have evolved from Bayesian Neural Networks and Monte Carlo Dropout to Deep Ensembles~\cite{Lakshminarayanan2017DeepEnsembles}. However, these methods require a separate dataset for training and incur high computational costs, limiting their direct use with VLMs~\cite{Gal2015DropoutBayesian, Gawlikowski2023UQSurvey}. We therefore evaluate open‑source VLMs for underwater trash detection using token-level probability‑based metrics (MSP, PCS, Deep Gini, Entropy)~\cite{Valle2025VLAUQ} and calibration metrics (ECE, MCE)~\cite{Kostumov2024VLMEvalUQ}.

\section{Conclusion and Future Work}\label{sec:conclusion}

Autonomous Underwater Robots (AURs) operate in challenging environments, performing tasks such as trash collection and environmental surveys. While deep learning has been used for these tasks, it is limited by its reliance on labeled data and inability to detect unseen objects, limitations that VLMs may overcome. This paper presents an empirical evaluation of VLM-based perception modules to assess their ability to detect underwater trash in AUR software. We evaluate four open-source VLMs (\blip, \deepseek, \lava, \qwen) on two datasets (\trashcan, \seaclear), analyzing performance, uncertainty, and their relationships across four classes: trash, object, animal, and vegetation. Results show that \blip performs the best overall, being the most calibrated and least overconfident, making it the most reliable for AUR software. In future work, we aim to integrate VLMs into AUR systems to detect unseen objects and develop testing approaches to ensure their dependable operation in marine tasks.

\section*{Acknowledgments}
This work is supported by the InnoGuard Doctoral Network under the Marie Skłodowska-Curie Actions of the European Commission, Grant Agreement No. 101169233. P. Arcaini is supported by the ASPIRE grant No. JPMJAP2301, JST.  Aitor Arrieta is part of the Software and Systems Engineering research group of Mondragon Unibertsitatea (IT1519-22), supported by the Department of Education, Universities and Research of the Basque Country.

\bibliographystyle{unsrtnat}
\bibliography{references}  

\end{document}